\documentclass[12pt]{article}
\usepackage{amsmath}
\usepackage{graphicx,psfrag,epsf}
\usepackage{enumerate}
\usepackage{natbib}
\usepackage{amsfonts} 
\usepackage{url} % not crucial - just used below for the URL 
\usepackage{hyperref}
\usepackage{graphicx}
\usepackage[english]{babel}
\usepackage{amsthm}
\usepackage{bbm}
\usepackage{soul}
\usepackage{paralist}
\usepackage{lipsum}
\usepackage{enumitem}
\usepackage{color}
\usepackage{adjustbox}
\usepackage{booktabs}
\setlist[enumerate]{itemsep=-2mm}
\setlist[itemize]{itemsep=-2mm}

\newtheorem{assumption}{Assumption}

\newtheorem{example}{Example}

\theoremstyle{definition}
\newtheorem{definition}{Definition}%[section]

\theoremstyle{remark}
\newtheorem*{remark}{Remark}

\newcommand{\E}{\mathbb{E}}

\newcommand{\Z}{\mathbf{Z}}
\newcommand{\A}{\mathbf{A}}
\newcommand{\X}{\mathbf{X}}

\newcommand{\z}{\mathbf{z}}

\newcommand{\pr}{\mathrm{Pr}}

\newcommand{\eqd}{\mathop{=}^{D}}
\newcommand{\indsim}{\mathop{\sim}^{ind}}

\newcommand{\blind}{0}

\begin{document}

\def\spacingset#1{\renewcommand{\baselinestretch}%
{#1}\small\normalsize} \spacingset{1}

%%%%%%%%%%%%%%%%%%%%%%%%%%%%%%%%%%%%%%%%%%%%%%%%%%%%%%%%%%%%%%%%%%%%%%%%%%%%%%

\if0\blind
{
  \title{\bf Degree of Interference: A General Framework For Causal Inference Under Interference}
  \author{Yuki Ohnishi
  % \thanks{ The authors gratefully acknowledge}\hspace{.2cm}
    \\
    Department of Biostatistics, Yale School of Public Health,\\
    Bikram Karmakar \\
    Department of Statistics, University of Wisconsin-Madison
    and \\
    Arman Sabbaghi \\
    Department of Statistics, Purdue University}
  \maketitle
} \fi

\if1\blind
{
  \bigskip
  \bigskip
  \bigskip
  \begin{center}
    {\LARGE\bf Degree of Interference: A General Framework For Causal Inference Under Interference}
\end{center}
  \medskip
} \fi

\bigskip

\begin{abstract}
One core assumption typically adopted for valid causal inference is that of no interference between experimental units, i.e., the outcome of an experimental unit is unaffected by the treatments assigned to other experimental units. This assumption can be violated in real-life experiments, which significantly complicates the task of causal inference. As the number of potential outcomes increases, it becomes challenging to disentangle direct treatment effects from ``spillover'' effects. Current methodologies are lacking, as they cannot handle arbitrary, unknown interference structures to permit inference on causal estimands. We present a general framework to address the limitations of existing approaches. Our framework is based on the new concept of the ``degree of interference'' (DoI). The DoI is a unit-level latent variable that captures the latent structure of interference. We also develop a data augmentation algorithm that adopts a blocked Gibbs sampler and Bayesian nonparametric methodology to perform inferences on the estimands under our framework. We illustrate the DoI concept and properties of our Bayesian methodology via extensive simulation studies and an analysis of a randomized experiment investigating the impact of a cash transfer program for which interference is a critical concern. Ultimately, our framework enables us to infer causal effects without strong structural assumptions on interference.
\end{abstract}

\noindent%

\section{Introduction}

%There are several common assumptions that have been considered fundamental for 
The two components of the Stable Unit Treatment Value Assumption (SUTVA) \citep{Rubin1974} are fundamental for causal inference for a treatment effect. 
The first component of SUTVA is that there are well-defined treatments for each experimental unit that yield well-defined outcomes, and the second is that no interference exists between units. The first assumption is typically satisfied by the design of an experiment, whereas the second relates to the complicated phenomenon of interference and may not be true in certain experiments. Interference is said to arise when a particular experimental unit's outcome is a function of both their assigned treatment as well as the treatments assigned to other experimental units. For example, interference can naturally arise in an experiment on the efficacy of a vaccine for a contagious disease, as one subject's health outcome can depend on whether their close contacts received the vaccine. The existence of interference complicates causal inferences because one must disentangle direct treatment effects from the spillover effects of the treatments assigned to other units.

This paper proposes a Bayesian semiparametric methodology for causal inference under interference. It is designed for situations where only partial, though potentially multiple, insights are available regarding the interference. The existing methods for causal inference under interference, as we discuss in \S\ref{sec_lit}, either assume the interference structure or assess the robustness of inference when it is misspecified. In the latter case, the inference is generally sensitive to misspecification.
In our proposed approach, a unit's interference structure is modeled as a latent function of the unit's features and other units' treatments. This latent function identifies the direct and spillover effect. We develop a Bayesian nonparametric prior for this latent function and a fast data augmentation algorithm that adopts a blocked Gibbs sampler to draw posterior inferences for the causal effects. The code implementing our method is available in the online supplement.

We call the unit-level latent functions ``degree of interference.'' One of our main contributions is to formalize the notion of the degree of interference. Using this notion, we develop a context-aware modeling strategy that flexibly reflects our knowledge about the interference structure. In the next section, we review existing methods and highlight our contributions in their context.

\section{Current Literature on Causal Inference Under Interference}\label{sec_lit}
The first foundational work towards causal inference under interference was by \citet{Hudgens2008}, who proposed a design-based approach to identify and estimate direct and spillover effects in a two-stage hierarchical experimental design. The first stage of the design randomly assigns clusters of units to one of two treatment assignment strategies, and then the second stage assigns treatments within the clusters based on the selected strategies. Their methods assume that the interference is (1) partial, i.e., operates only within clusters, and (2) stratified, i.e., is specified via the proportion of treated units within clusters. Several important works have built upon the above assumptions \citep{Liu2014,Kang2016, sussman2017, Basse_avi2018, Imai2021, ohnishi2021bayesian}. However, in practice, the validity of the assumptions of partial interference, which requires multiple independent clusters, and stratified interference, which is typically a simplification for the unknown interference structure, may be questionable \citep[p.~836]{Hudgens2008}. When these assumptions fail, the above methodologies provide biased inferences \citep{Fredrik2021_2}. 
In contrast,  we utilize a Bayesian nonparametric prior for the ``degree of interference'', specifically, the Dependent Dirichlet process mixture prior to flexibly model the unknown interference structure. These priors can data-adaptively separate units into several clusters based on how they interfere with one another, thus accommodating a stratified nature of the interference structure.

\citet{Aronow2017} introduced the idea of exposure mapping to move beyond the partial and stratified interference assumptions. Under this framework, a mapping is specified that relates the vector of treatment assignments for the other experimental units to a finite set of exposures. Then, an experimental unit's outcome is specified as a function of its own treatment label and this exposure. In other words, an exposure map is supposed to capture the full nature of interference of a unit from all other units. Once an exposure map is well-specified, causal effects are defined in terms of comparisons of outcomes under different exposures. A line of research exists that uses exposure maps to provide causal inference under interference. 
Notably, \citet{Toulis2013} developed randomization-based and Bayesian model-based inferential approaches for ``peer influence effects'' defined by fixing the specific number of neighbors who receive treatment. 
\citet{Basse2019} developed a framework for constructing valid randomization tests for general exposure contrasts. 
Other research works have addressed interference by specifying their interference mapping based on peers' treatment status and attributes of a given network \citep{Xinran2019, Zigler2021, Forastiere2021, papadogeorgou_Samanta2023}.% These approaches can be viewed as a specific instance of exposure mapping.}

As \citet{Basse_Airoldi2018} mention, causal inference under interference is impossible without assumptions about the interference structure. Thus, it is natural to use exposure maps specifying the structure. However, current causal inferential frameworks and methodologies based on exposure maps are limited in several respects. 
First, they require that the interference structure be fully characterized as a finite set of exposures \emph{a priori} so that each exposure corresponds to a unique potential outcome. In practice, satisfying this requirement may be unrealistic because the interference structure is usually unknown, or partially known, and can be very complicated to understand.  Second, as observed by \citet{Fredrik2021_2}, a misspecified exposure mapping can lead to imprecise variance estimation and, thus, incorrect inference. 
The domain knowledge of the interference structure can mitigate this problem. \citet{Leung2021} proposed a flexible model that permits treatments that are assigned to units ``farther'' from a particular unit to have non-vanishing, but smaller, effects on the unit's response. Our proposal for the ``degree of interference'' concept generalizes the exposure map. Unlike other methods, it does not require precise knowledge of the interference structure.  We specify a flexible semi-parametric model for the interference structure and define the potential outcomes as expected values of pseudo-potential outcomes that marginalize with respect to the uncertainty in the interference model. This approach allows for knowledge-aware specifications about interference and a fast Bayesian algorithm to infer the counterfactual outcomes.
%that enable us to incorporate whatever knowledge we have about the interference structure into our inferences.

Third, the exposure mapping framework simultaneously defines both the interference structure that determines a single unique potential outcome and exposure effects that determine causal quantities. However, since exposure mapping is typically problem-specific, exposure effects generally do not hold any meaning outside of the problem instance, even for the same treatment-outcome pair. Recent literature building upon the exposure mapping framework aims to unbiasedly estimate the exposure effects (e.g., \cite{Fredrik2021_2, Leung2021}) or to construct the testing procedure for the correctness of the exposure map \cite{hoshino2023}. In contrast, we define new causal estimands that are not bonded with the interference structure. Our estimands are generalizations of those suggested by \cite{Fredrik2021} as natural causal estimands for direct and spillover effects under interference. 
\cite{Fredrik2021} demonstrated that the Horvitz-Thompson (HT) and H\"ajek estimators are consistent and derived their asymptotic distributions. However, these estimators lack theoretical guarantees when interference scales proportionally with the sample size, as outlined in Assumption 2 of \citet{Fredrik2021}. Our framework circumvents such restrictions by allowing us to model the interference structure and serves as a promising alternative in such a complex interference setting.
Finally, our approach allows for continuous treatment factors and factors with many discrete levels. This flexibility is particularly advantageous in many practical settings, where such variations in treatment factors are commonplace.

The rest of the paper is organized as follows. We proceed in \S\ref{sec:setup} to review the Rubin Causal Model and propose new spillover effect estimands under interference. \S\ref{sec:framework} contains our new framework based on the ``degree of interference'' concept and our corresponding data augmentation algorithm for inferring causal estimands under interference. The frequentist properties of our method are demonstrated via extensive simulation studies in \S\ref{sec:simulation_studies}. In \S\ref{sec:case_studies_main} we demonstrate our framework using case studies on cash transfer programs in Colombia and an experiment on stereolithography. 
\S\ref{sec:conclusions} concludes the paper.% with a discussion on future research.

\section{Background}
\label{sec:setup}

\subsection{The Rubin Causal Model and Interference Structures}
\label{sec:RCM}

%We adopt the Rubin Causal Model as our causal paradigm where the first step is to define the Science of a setting, i.e., to define the experimental units, covariates, treatments, and potential outcomes \citep{Rubin1974, imbens_rubin_2015}. 
Adopting the Rubin Causal Model \citep{Rubin1974},
consider $N$ experimental units, indexed by $i = 1, \ldots , N$, that correspond to physical objects at a particular point in time. Each experimental unit $i$ has an observed set of pre-treatment covariates $\mathbf{X}_i \in \mathcal{X}$, with $\mathcal{X}$ denoting the covariate space; typically; $\mathcal{X} = \mathbb{R}^d$ for $d$ covariates. Treatment assignments are contained in the $N$-dimensional vector $\mathbf{Z} = (Z_1, \ldots, Z_N)^\top$, where $Z_i \in \Omega$, the treatment assignment for unit $i$, can be binary, categorial or continuous, and $\Omega$ is the set of treatment levels. So, the vector $\mathbf{Z}\in \Omega^N$. %We do not restrict the $Z_i$ to be binary treatments, and they could have more than two levels as either categorical or continuous treatment factors. 
For example, $\Omega = \mathbb{R}$ denotes a continuous treatment factor with ``$0$'' denoting the specified ``control'' level, and $\Omega = \{0, 1, \ldots, K\}$ denotes a categorical treatment with $K+1$ levels and with ``$0$'' again denoting control. Let $\mathbf{Z}_{-i} = (Z_1, \ldots, Z_{i-1}, Z_{i+1}, \ldots, Z_N)^\top$ be the sub-vector of $\mathbf{Z}$ with the $i$th element removed; so, $\mathbf{Z}_{-i}\in \Omega^{N-1}$. 

Next, consider the potential outcomes for unit $i$. To facilitate the exposition and indicate how, in the general setting of interference, unit $i$'s potential outcomes depend on multiple entries in $\mathbf{Z}$, we write the potential outcomes for unit $i$ as $Y_i(\mathbf{Z}) \equiv Y_i(Z_i, \mathbf{Z}_{-i})$. We let $\mathbf{Y} = \{ Y_i(\mathbf{z}): i = 1, \ldots, N, \mathbf{z} \in \Omega^N \}$ denote the set of all potential outcomes. The observed outcomes are contained in vector $\mathbf{Y}^{\mathrm{obs}} = (Y_1^{\mathrm{obs}}, \ldots, Y_N^{\mathrm{obs}})^\top$ where $Y_i^{\mathrm{obs}} = \sum_{\mathbf{z} \in \Omega^N} Y_i({\mathbf{z}})\mathbb{I}(\mathbf{Z} = \mathbf{z})$ for each $i$. %Superscript ``$\mathrm{obs}$'' denotes an observed realization. 
Thus, only one outcome is observable to us among the many potential outcomes of unit $i$.

The probability distribution of $\mathbf{Z}$ given covariates $\mathbf{X}_1, \ldots, \mathbf{X}_N$ is known to us. 
We assume without loss of generality that 
the support of the probability distribution is equal to $\Omega^N$.% without loss of generality.

Since the experimental units interfere, as is common in the literature, we assume throughout that they belong to a single, known network that governs their interference structure. Let $\mathbf{A}$ denote the $N \times N$ matrix that captures how the units are related to one another. The specification of $\mathbf{A}$ can vary depending on how one assumes interference between units to occur. For example, to describe the interference structure based just on unit adjacencies, $\mathbf{A}$ can be defined simply via the adjacency matrix of the units. 
If instead one wishes to use the shortest path between units to describe their interference structure, then $\mathbf{A}$ is defined as the distance matrix of the units, with element $(i,j)$ of $\mathbf{A}$ being the shortest distance between units $i$ and $j$. Later, in our definition of the degree of interference, $\mathbf{A}$ becomes an argument of its functional form.

% \subsection{Dirichlet process mixture}

\subsection{Causal Estimands}
\label{sec:causal_estimands}

% \subsubsection{Types of Causal Estimands Under Interference}
Throughout, we consider the finite-population perspective, where we view the $N$ units are given, and an external, known randomization process assigns treatment to these units. All potential outcomes are defined as fixed but mostly missing for the unrealized treatment assignment. 

We define two types of causal estimands---the first type is assignment-conditional effects, and the second is expected effects marginalized over a treatment assignment. Assignment-conditional causal estimands are defined by conditioning on a specific treatment assignment vector. We consider two effects for this type of estimand: average treatment and spillover effects.

%\label{sec:ate}
%\subsubsection{Assignment-conditional Treatment Effects}

Following \citet{Fredrik2021}, who used similar estimands in the case of a binary treatment, the unit-level assignment-conditional treatment effect for unit $i$ is the contrast between its two potential outcomes corresponding to $Z_i = 0$ and $Z_i = z$ when the treatment assignments $\mathbf{Z}_{-i}$ for the other experimental units are held fixed at $\z_{N-1} \in \Omega^{N-1}$. Denote this effect by $\tau_i^{\mathrm{tr}}(Z_i=z, \mathbf{Z}_{-i} = \z_{N-1}) = Y_i(Z_i = z, \mathbf{Z}_{-i} = \z_{N-1}) - Y_i(Z_i = 0, \mathbf{Z}_{-i} = \z_{N-1})$. The average of the $\tau_i^{\mathrm{tr}}(Z_i=z, \Z_{-i} = \z_{N-1})$ across the $N$ experimental units corresponds to a finite-population average treatment effect.

%that is well-defined under interference.

\begin{definition}[Conditional Average Treatment Effect]
\label{def:A-CATE}
An \underline{a}ssignment-\underline{c}onditional \underline{a}verage \underline{t}reatment \underline{e}ffect (A-CATE) comparing a pair of treatment, a $z \in \Omega$ and vector $\z' \in \Omega^{N}$ is
%is the average of the unit-level assignment-conditional treatment effects under a treatment $z \in \Omega$ for unit $i$ and a treatment vector $\z' \in \Omega^{N}$ for the other units: 
$$\tau_{\text{A-CATE}}(z, \z') = \frac{1}{N}\sum_{i=1}^{N} \tau_i^{\mathrm{tr}}(Z_i = z, \mathbf{Z}_{-i} = \z_{-i}'),$$ 
where $\z_{-i}'$ is an $(N-1)$-dimensional sub-vector of $\z'$ with element $i$ removed.
%\begin{equation}
%\label{eq:A_CATE}
%\begin{split}
%    \tau_{\text{A-CATE}}(z, \mathbf{z}_{-i}) &= \frac{1}{N} \sum_{i=1}^{N} \tau_i^{\mathrm{tr}}(Z_i = z, \mathbf{Z}_{-i} = \z_{-i}) \\
%&= \frac{1}{N} \sum_{i=1}^{N} \left \{ Y_i(Z_i = z, \mathbf{Z}_{-i} = \mathbf{z}_{-i}) - Y_i(Z_i = 0, \mathbf{Z}_{-i} = \mathbf{z}_{-i}) \right \} .
%\end{split}
%\end{equation}
\end{definition}
\noindent A-CATE is a natural extension of the estimand of \citet{Fredrik2021} for the cases of categorical or continuous treatments. This effect quantifies the average change in the outcome as one unit's treatment changes from $0$ to $z$ when the treatments for all other units are fixed at $\mathbf{z}_{-i}'$.

We define our unit-level spillover effect that is inspired by the A-CATE. Specifically, for unit $i$ with treatment $z$ and for two treatment assignment vectors $\mathbf{z}_{N-1}', \mathbf{z}_{N-1}^{*} \in \Omega^{N-1}$ for the other units, the unit-level spillover effect is defined as $\tau_i^{\mathrm{sp}}(z, \z_{N-1}', \z_{N-1}^{*}) = Y_i(Z_i = z, \mathbf{Z}_{-i} = \mathbf{z}_{N-1}') - Y_i(Z_i = z, \mathbf{Z}_{-i} = \mathbf{z}_{N-1}^{*})$.

\begin{definition}%[Assignment-Conditional Average Spillover Effect]
\label{def:A-CASE}
An \underline{a}ssignment-\underline{c}onditional \underline{a}verage \underline{s}pillover \underline{e}ffect (A-CASE) under a treatment $z \in \Omega$ for unit $i$ and treatment vectors $\mathbf{z}_{N}', \mathbf{z}_{N}^{*} \in \Omega^{N}$
%is the average of the unit-level assignment-conditional spillover effects under a treatment $z \in \Omega$ for unit $i$ and treatment vectors $\mathbf{z}_{N}', \mathbf{z}_{N}^{*} \in \Omega^{N}$ for the other units: 
$$\tau_{\text{A-CASE}}(z, \mathbf{z}_{N}', \mathbf{z}_{N}^{*}) = \frac{1}{N}\sum_{i=i}^{N} \tau_i^{\mathrm{sp}}(z, \mathbf{z}_{-i}', \mathbf{z}_{-i}^{*}),$$ where $\z_{-i}'$ and $\z_{-i}^{*}$ are $(N-1)$-dimensional sub-vectors of $\z'$ and $\z^{*}$ with the $i$th elements removed.
%\begin{align*}
%\tau_{\text{A-CASE}}(z, \mathbf{z}_{-i}, \mathbf{z}_{-i}') &= \frac{1}{N} \sum_{i=i}^{N} \tau_i^{\mathrm{sp}}(z, \mathbf{z}_{-i}, \mathbf{z}_{-i}') \\
%&= \frac{1}{N} \sum_{i=i}^{N} \left \{ Y_i(Z_i = z, \mathbf{Z}_{-i} = \mathbf{z}_{-i}) - Y_i(Z_i = z, \mathbf{Z}_{-i} = \mathbf{z}_{-i}') \right \} .
%\end{align*}
\end{definition}

\noindent Our definition of the A-CASE corresponds to the average comparison of the units' outcomes for a single treatment under two different cases, corresponding to the two different treatment vectors considered for the other units. Of special interest is the case with $z = 0, \mathbf{z}_{N}^{*} = (0, \ldots, 0)^\top$, and $\mathbf{z}'$ set at some other vector, where A-CASE captures a pure spillover effect.

Our second type of causal estimands is a set of three expected effects, which marginalize the assignment-conditional causal estimands over a probability distribution on the assignment vectors, i.e., the treatment assignment mechanism. Expected effects are useful in practice because the number of assignment-conditional effects grows rapidly as a function of $N$, whereas the number of expected effects is generally smaller. 
% (as they are defined conditional on the assignment vector, and the number of possible vectors increases exponentially as a function of $N$)
% It thus becomes easier to use and expected effects instead of assignment-conditional effects when performing causal inference under interference.

\begin{definition}[Expected Average Treatment and Spillover]
\label{def:expected_effects}
Let $\mathbb{E}_{\pi}(h(\Z))$ denote the expectation of a function $h(\Z)$ of the assignment vector over the known probability mass/density function $\pi(\mathbf{z}) = p(\mathbf{z} \mid  \mathbf{X}_1, \ldots, \mathbf{X}_N)$. Then for each assignment-conditional average effect, we define the corresponding expected average effect based on the expectation of the assignment-conditional average effect over the assignment mechanism: $\tau_{\text{E-ATE}}(z;\pi) = \mathbb{E}_{\pi}\{ \tau_{\text{A-CATE}}(z,\mathbf{Z}) \}$ and $ \tau_{\text{E-ASE}}(z;\pi) = \mathbb{E}_{\pi}\{ \tau_{\text{A-CASE}}(z, \mathbf{Z},\mathbf{0}) \}$. 
% and $\tau_{\text{E-AOE}}(\pi) = \mathbb{E}_{\pi}\{ \tau_{\text{A-CAOE}}(\mathbf{Z}) \}$. 
\end{definition}
\noindent
Our version of this estimand, $\tau_{\text{E-ATE}}$, is a natural extension of the estimand of \citet{Fredrik2021} for categorical treatment factors with multiple levels and continuous factors. Our definitions of $\tau_{\text{E-ASE}}$ is motivated in a similar manner as that of $\tau_{\text{E-ATE}}$. As noted by \citet{Fredrik2021}, the E-ATE---similarly, E-ASE---captures the expected average effect of changing the treatment of a single unit in the ``current'' experiment. %The E-ASE and E-AOE also capture the expected average effect of changing the treatment vectors for all units in the current experiment. 

\section{The Degree of Interference Framework and Bayesian Methodology}
\label{sec:framework}

\subsection{Definition and Assumptions for the Degree of Interference }
\label{sec:DOI}

The Degree of Interference (DoI) is defined in terms of the characteristics of a unit and the treatments for all other units. 
Our DoI framework addresses a limitation of exposure maps that, while an exposure map needs to be pre-specified, the DoI is an unknown latent function that represents the underlying interference structure between units, which is inferred from the data.

We let characteristics $\mathbf{T}_i \in \mathcal{T}$ for unit $i$ contain, its covariates, $\mathbf{X}_i$, covariates for units other than $i$, $\mathbf{X}_{-i}$, and the adjacent network information of the other units. In all that follows, we let $\mathcal{T} = \mathcal{X} \: \times \: \mathcal{X}^{N-1} \times \: \mathcal{A}$, where $\mathcal{A}$ denotes the set of matrices of networks, i.e., set of $\mathbf{A}$s.

\begin{definition}
\label{def:DoI}
The DoI for unit $i$ is a random function $G_i:\mathcal{T} \times \Omega^{N-1} \to \mathbb{R}$ that maps the unit's characteristics, residing in $\mathcal{T}$, and treatments for all other units to a number. 
\end{definition}

%{\color{red} BK: when talking about the no covariates scenario, we should also note the mathematical form of the function  $G_i$. $\mu$ is not defined or mentioned up to this point. We need to clarify this.}

\noindent

In a simple scenario with no individual or subgroup covariates, each unit’s ``characteristics'' only encode the network or grouping structure that dictates which other units can affect it. In this setting, $G_i$ is a stochastic process indexed by an element $(\mathbf{A}, \Z_{-i}) \in \mathcal{A} \times \Omega^{N-1}$.
For example, if it is reasonable to assume that the proportion of treated units in a neighborhood is the primary factor determining the spillover effect (i.e., the stratified interference assumption; \citep{Hudgens2008}), we may assume $G_i \sim \mathrm{N}\!\left(\beta_0 + \beta_1 \frac{\sum_{j \in \mathcal{N}_i} Z_j A_{ij}}{|\mathcal{N}_i|}, \sigma^2\right),$
where $\mathcal{N}_i$ is the set of neighboring units for unit $i$.
The core of the DoI framework allows the function $G_i$---and the corresponding parameters $\beta_0$, $\beta_1$, and $\sigma$---to be unknown, latent quantities inferred from the data rather than a fixed, pre-specified exposure map. For instance, one might be confident that the ``proportion of treated neighbors'' is relevant for interference but remain unsure whether it is indeed the correct functional form.
In practice, the DoI framework can learn such dependencies from data, even if there is a more complex network relationship (e.g., higher-order neighbors). By contrast, the \emph{exposure map} approach would require the analyst to explicitly fix a certain functional form (like a fraction of treated neighbors), potentially overlooking more nuanced patterns.
Detailed discussions are provided in Section \ref{sec:structural_assumption}.

% \noindent 
The DoI satisfies the following assumption which says that unit $i$'s potential outcomes are identified from its own treatment and DoI.
\begin{assumption}
\label{asmp:doi}
For all $i \in \{1,...,N\}$, $\mathbf{T}_i \in \mathcal{T}$, $z \in \Omega$, and $\z_{N-1},\z_{N-1}' \in \Omega^{N-1}$ in which $G_i(\mathbf{T}_i, \z_{N-1}) \: \displaystyle{\eqd} \: G_i(\mathbf{T}_i, \z_{N-1}')$, where $\displaystyle{\eqd}$ denotes equivalence in the marginal prior distributions of $G_i(\mathbf{T}_i, \z_{N-1})$ and $G_i(\mathbf{T}_i, \z_{N-1}')$, we have
$Y_i(z,\z_{N-1})  = Y_i(z,\z_{N-1}')$.
\end{assumption}
\noindent
This assumption ultimately corresponds to equivalence in the potential outcomes according to DoIs that have the same marginal prior distributions. More formally, for all $i \in \{1,...,N\}, z \in \Omega, \z_{N-1} \in \Omega^{N-1}$, and $\mathbf{T}_i \in \mathcal{T}$, the equivalencies of the potential outcomes for unit $i$ with $ Y_i(z,\z_{N-1})$ are based on $z$ and the marginal prior distribution on 
$G_i \equiv G(\mathbf{T}_i, \z_{N-1})$.
We represent its conditional cumulative distribution function 
$F_{G(\mathbf{T}_i, \z_{N-1})}$. 

\begin{remark}[Randomness of DoI]
Assumption \ref{asmp:doi} is critical, as it implies that the DoI is effectively a latent representation of the interference structure for unit $i$ with characteristics $T_i$ when other units are assigned to $\z_{N-1}$.
The distributional equivalence allows for additional flexibility of the DoI; it indicates the unknown uncertainty about the interference structure. Under the finite-population perspective, the treatment assignment is usually the only source of randomness for observables. Therefore, the existing design-based approach under the finite-population perspective (e.g., partial interference, stratified interference and exposure maps) inevitably defines the interference structure \emph{a priori} even when there is limited knowledge about it. 
Under our framework, by contrast, the interference structure is an ``unobserved'' quantity, representing uncertainty for analysts. 
We incorporate the knowledge about interference and the associated uncertainty into our model-based inference developed in the following sections. 
% More specifically, we use a  Bayesian approach to infer unknown DoIs, requiring us to specify their prior distributions. 
% From the Bayesian perspective, the randomness of the DoI can be viewed as the representation of prior knowledge and uncertainty about the interference structure. 
Our DoI framework adopts a Bayesian perspective by introducing a random function $G_i$ to capture how other units' treatments and network structures affect unit $i$. Although $Y_i(\mathbf{z})$ is ultimately fixed, from the analyst's point of view it remains \emph{unknown} and is therefore modeled through a prior distribution over $G_i$. This approach allows us to incorporate uncertainty about the interference mechanism and update our beliefs based on observed data. In other words, the DoI framework posits a prior belief about how cross-unit effects flow through the latent function $G_i$, and subsequent inferences reflect how the data inform and refine our understanding of this interference channel.
This uncertainty should be distinguished from the ones on the underlying network between units \citep{Egami2021}.
We place priors on functional spaces, which are flexible enough to recover the latent DoIs that satisfy Assumption \ref{asmp:doi}.  
\end{remark}

In addition to Assumption \ref{asmp:doi}, we consider the following assumption about the relation between the DoI function and potential outcomes.
\begin{assumption}
\label{asmp:aux_po}
    For each unit $i$, there exists auxiliary potential outcomes $\tilde{Y}_i(z, g)$, where $\tilde{Y}_i(z, G_i)$ and $\tilde{Y}_i(z, G'_i)$ have the same distribution when $G_i$ and $G'_i$ have the same marginal prior distributions. In contrast to the original potential outcome, the auxiliary potential outcomes have randomness induced by  $G_i$, and the original potential outcome is:
    \begin{equation}
    \label{eq:aux_po}
    Y_i(z,\z_{N-1}) = \E_{G(\mathbf{T}_i, \z_{N-1})} \{\tilde{Y}_i(z,G_i) \}=
    \int_{\mathbb{R}} \tilde{Y}_i(z,g) dF_{G_i(\mathbf{T}_i, \z_{N-1})} (g),
    \end{equation}
\end{assumption}

\noindent
Under this assumption, our DoI framework naturally incorporates the uncertainty associated with interference dynamics into the process of inferring potential outcomes. In contrast, the exposure mapping framework assumes exact specification of the interference structure and thus is limiting. Further, an exact specification of the interference is often not possible in practice. Also, our framework under Assumptions \ref{asmp:doi} and \ref{asmp:aux_po} generalizes the exposure mapping framework since our framework does not assume a particular functional form for $F_{G_i(\mathbf{T}_i, \z_{N-1})}$ beforehand, as the exposure mapping does.

Equation \eqref{eq:aux_po} says that the uncertainty about $G_i$ is incorporated by marginalizing the auxiliary potential outcomes over $G(\mathbf{T}_i, \z_{N-1})$ for the original potential outcomes with $\Z_{-i}=\z_{N-1}$. Also, it implies an inferential approach for $Y_i(z,\z_{N-1})$ via the Monte Carlo simulation: $Y_i(z,\z_{N-1}) \approx
\sum_{m=1}^{M} \tilde{Y}_i(z,g_m) / M $ where $g_m \sim F_{G_i(\mathbf{T}_i, \z_{N-1})}$ for $m=1,\ldots,M$. The Monte Carlo procedure is introduced in  \S\ref{sec:Gibbssampling_outline}.
Finally, the unit-level assignment-conditional treatment effect is
\begin{equation}
\label{eq:eff_tr_g}
\begin{split}
&\tau_i^{\mathrm{tr}}(Z_i=z, \Z_{-i}=\z_{N-1})= \int_{\mathbb{R}}   \big\{\tilde{Y}_i(z, g) - \tilde{Y}_i(0, g)\big\}  dF_{G_i(\mathbf{T}_i, \z_{N-1})}(g).
\end{split}
\end{equation}

% This assumption ultimately corresponds to equivalence in the potential outcomes according to DoIs that have the same marginal prior distributions. More formally, for all $i \in \{1,...,N\}, z \in \Omega, \z_{N-1} \in \Omega^{N-1}$, and $\mathbf{T}_i \in \mathcal{T}$, the equivalencies of the potential outcomes for unit $i$ with $ Y_i(z,\z_{N-1})$ are based on $z$ and the marginal prior distribution on 
% $g_i \equiv G_i(\mathbf{T}_i, \z_{N-1})$.
% We represent its conditional cumulative distribution function 
% $F_{G(\mathbf{T}_i, \z_{N-1})}$. 
% Hence, as a consequence of Assumption \ref{asmp:doi}, we use potential outcomes $\tilde{Y}_i(z, g)$, where $\tilde{Y}_i(z, g_i) =  \tilde{Y}_i(z, g'_i) $ for any $g_i, g'_i$ that have the same marginal prior distributions. Then we have that
% \begin{align*}
% Y_i(z,\z_{N-1}) &= Y_i(z,\z_{N-1}) \int_{\mathbb{R}} dF_{G_i(\mathbf{T}_i, \z_{N-1})} (g_i)\\
% &= 
% \int_{\mathbb{R}} Y_i(z,\z_{N-1}) dF_{G_i(\mathbf{T}_i, \z_{N-1})} (g_i) = 
% \int_{\mathbb{R}} \tilde{Y}_i(z,g) dF_{G_i(\mathbf{T}_i, \z_{N-1})} (g_i),
% \end{align*}
% where the last equality follows by Assumption \ref{asmp:doi}. Thus, the unit-level assignment-conditional treatment effect is
% \begin{equation}
% \label{eq:eff_tr_g}
% \begin{split}
% &\tau_i^{\mathrm{tr}}(Z_i=z, \Z_{-i}=\z_{N-1})= \int_{\mathbb{R}}   \big\{\tilde{Y}_i(z, g) - \tilde{Y}_i(0, g)\big\}  dF_{G_i(\mathbf{T}_i, \z_{N-1})}(g_i).
% \end{split}
% \end{equation}
% \yo{\bf remove above?}

\noindent Unit-level assignment conditional spillover effects can also be expressed in terms of the distribution of the DoI. Equation \eqref{eq:eff_tr_g} further clarifies the conceptual role of the DoI that it is effectively a low-dimensional compression of interference, and the direct effect is defined as the comparison of two auxiliary potential outcomes with different treatment assignments under the same DoI.

DoI, along with Assumptions \ref{asmp:doi} and \ref{asmp:aux_po}, generalizes exposure mappings \citep{Aronow2017}. In particular, if  $F_{G_i(\mathbf{T}_i, \z_{N-1})}$ is a prespecified, known function that maps to a finite set then the DoI is analogous to exposure mappings. For example, the exposure mapping for stratified interference \citep{Hudgens2008} would correspond to prespecifying $F_{G_i(\mathbf{T}_i, \z_{N-1})}$ as an indicator function of the number of treated units in the neighborhood of unit $i$:
$F_{G_i(\mathbf{T}_i, \z_{N-1})}(g) = \mathbbm{1} (\z_{N-1}^\top A_{-i,i}=g)$ for $g \in \{1, \ldots, N-1\}$,
where $A_{-i,i}$ is the $i$-th column vector of the adjacency matrix $A$ with the $i$-th row removed.
However, in contrast to existing approaches based on exposure maps, the DoI is not prespecified but is instead inferred from data via our Bayesian methodology in \S\ref{sec:methodology_overview} and \S\ref{sec:models}. 
%It is unlikely that one could precisely prespecify the interference structure in a particular real-life setting. As such, inferring the DoI is more relevant compared to existing approaches, e.g., imposing a stratified interference structure on a setting from the start when it could in fact be a coarse approximation of interference and likely fail to capture it. 
Inferring the DoI is advantageous in principle, as imposing a misspecified interference structure and failing to modify it could yield biased inferences \citep{Fredrik2021_2}.

Finally, the DoI framework allows causal inferences for experiments that satisfy the following new version of the latent unconfoundedness assumption.
\begin{assumption}
\label{asmp:unconfoundedness}
A treatment assignment is unconfounded under the DoI framework if the probability mass/density function in the treatment assignment mechanism does not depend on $\mathbf{Y}$ or the $G_i(\mathbf{T}_i, \mathbf{z}_{-i})$ conditional on the $\mathbf{T}_i$, i.e., $p(\mathbf{z} | \mathbf{T}_1, \ldots, \mathbf{T}_N, \mathbf{Y}, G_1(\mathbf{T}_1, \mathbf{z}_{-1}'), \ldots,  G_N(\mathbf{T}_N, \mathbf{z}_{-N}')) = p(\mathbf{z} | \mathbf{T}_1, \ldots, \mathbf{T}_N)$, for all $\mathbf{T}_1, \ldots, \mathbf{T}_N \in \mathcal{T}$, $\mathbf{z}, \mathbf{z}' \in \Omega^N$ and $G_1(\mathbf{T}_1, \mathbf{z}_{-1}), \ldots, G_N(\mathbf{T}_N, \mathbf{z}_{-N})$.
\end{assumption}
\noindent While this assumption is less restrictive than a randomized treatment assignment, this assumption may not be valid in observational studies. In our case studies in \S\ref{sec:case_studies_main}, Assumption \ref{asmp:unconfoundedness} is valid due to the design of the experiment.

\subsection{Overview of the Bayesian Methodology}
\label{sec:methodology_overview}

Bayesian inference is one of the modes of inference within the potential outcome framework, where we consider the missing potential outcomes as unobserved random variables, no different from unknown parameters. Specifically, Bayesian inference involves specifying a model for all random variables, including the potential outcomes, treatment, and covariates. Based on this model, one can infer causal estimates from the joint posterior distributions of the parameters and the missing potential outcomes, conditional on the observed data \citep{Rubin1978}. 
%The Bayesian paradigm provides a unified inferential framework for any causal estimand with automatic uncertainty quantification that does not rely on large-sample approximation. 

The Bayesian approach not only provides a principled framework for analyzing causal studies but also provides a refined map of identifiability, clarifying what can be learned when causal estimates are not fully identifiable but instead weakly identifiable (i.e., when the likelihood functions of parameters and causal estimates have substantial regions of flatness) \citep{ImbensRubin1997}.
In particular, issues of identification differ from those in the frequentist paradigm because if prior distributions are proper, then posterior distributions are always proper. Weak identifiability is reflected in the flatness of the posterior distribution and can be evaluated quantitatively \citep{Gustafson2009}.
\citet{FanLi2022} gives further comprehensive reviews of Bayesian causal inference. 

Our Bayesian methodology involves multiple imputations of the DoI for missing potential outcomes to derive the posterior distributions of the finite-population causal estimands. To formally describe this, let $\tau$ denote one of the causal estimands from \S\ref{sec:causal_estimands}, $\mathbf{T}$ the $N \times P$ matrix of characteristics for all units (where the columns correspond to the concatenation of covariates and adjacent network information), $\mathbf{Z}$ the $N \times 1$ vector of assigned treatments, $\mathbf{Y}^{\mathrm{obs}}$ the vector of observed outcomes, and $\mathbf{Y}^{\mathrm{mis}}$ the vector of all missing potential outcomes. The $\mathbf{Y}^{\mathrm{mis}}$ is determined according to the treatment assignment, with each missing potential outcome for unit $i$ being either $Y_i(z_i', \z_{-i})$ or $Y_i(z, \z_{-i}')$ with $z_i'$ denoting an alternate treatment for unit $i$ and $\z_{-i}'$ denoting an alternate treatment vector for all units excluding unit $i$. In these cases, the DoI for unit $i$ would be either $G_i(\mathbf{T}_i, \z_{-i})$ or $G_i(\mathbf{T}_i, \z_{-i}')$ respectively. We let $\mathbf{G}^{\mathrm{o}} = (G_1(\mathbf{T}_1, \mathbf{Z}_{-1}), \ldots, G_N(\mathbf{T}_N, \mathbf{Z}_{-N}))^\top$ and $\mathbf{G}^{\mathrm{u}}(\mathbf{z}') = (G_1(\mathbf{T}_1, \z_{-1}'), \ldots, G_N(\mathbf{T}_N, \z_{-N}'))^\top$ contain the DoIs for the realized and an unrealized treatment assignment vector $\mathbf{z}'$, respectively. None of the entries in $\mathbf{G}^{\mathrm{o}}$ or the $\mathbf{G}^{\mathrm{u}}(\mathbf{z}')$ are observable. In accordance with the matrix $\mathbf{Y}^{\mathrm{mis}}$, we generally define the matrix $\mathbf{G}^{\mathrm{u}}$ as the column-wise concatenation of all the $\mathbf{G}^{\mathrm{u}}(\mathbf{z}')$ for treatment assignment vectors $\mathbf{z}'$ that were not realized.

%We represent the posterior distribution of the estimand conditional on all observed data by the probability density function . 
Standard types of point estimates of estimand $\tau$ are obtained via means, medians, or modes of the posterior distribution $p \left ( \tau \mid \mathbf{T}, \mathbf{Z}, \mathbf{Y}^{\mathrm{obs}} \right )$, and standard interval estimates are obtained via the central credible intervals or highest posterior density intervals. As causal estimands are functions of observed and missing potential outcomes, we recognize that we can calculate this posterior distribution by integrating out $\mathbf{Y}^{\mathrm{mis}}, \mathbf{G}^{\mathrm{o}}$, and $\mathbf{G}^{\mathrm{u}}$ according to
\begin{equation}
 \int  p \left ( \tau \mid \mathbf{T},  \mathbf{Z}, \mathbf{Y}^{\mathrm{obs}}, \mathbf{G}^{\mathrm{o}}, \mathbf{G}^{\mathrm{u}}, \mathbf{Y}^{\mathrm{mis}} \right ) 
  p \left ( \mathbf{G}^{\mathrm{o}}, \mathbf{G}^{\mathrm{u}}, \mathbf{Y}^{\mathrm{mis}} \mid \mathbf{T},  \mathbf{Z}, \mathbf{Y}^{\mathrm{obs}} \right ) d\mathbf{G}^{\mathrm{o}} d\mathbf{G}^{\mathrm{u}} d\mathbf{Y}^{\mathrm{mis}}.
\label{eq:posterior_of_estimand}
\end{equation}
\noindent In order to sample from this posterior, we first derive the posterior distribution of the DoI and missing outcomes conditional on the observed data. Then, we calculate the posterior distribution of the estimand conditional on the observed data and imputations of the DoI and missing outcomes drawn from their posterior distributions. We perform all imputations and posterior calculations via Markov Chain Monte Carlo technique.  

Four inputs are necessary to derive the posterior distributions of the causal estimands. (i) The first is knowledge of the assignment mechanism, as encoded in the probability mass/density function $p \left ( \mathbf{Z} \mid \mathbf{T} \right )$.
% (for this methodology we only consider studies that satisfy Assumption \ref{asmp:unconfoundedness}). 
As we consider designed experiments, this probability mass/density function is known. 
(ii) The second is the prior distribution for the DoIs conditional on $\mathbf{T}$ and $\mathbf{Z}$, which we denote by the joint probability density function $p \left ( \mathbf{G}^{\mathrm{o}}, \mathbf{G}^{\mathrm{u}} \mid \mathbf{T}, \mathbf{Z}, \boldsymbol{\phi} \right )$ with a parameter vector $\boldsymbol{\phi}$. 
(iii) The third input is a model for the potential outcomes conditional on $\mathbf{T}, \mathbf{Z}, \mathbf{G}^{\mathrm{o}}, \mathbf{G}^{\mathrm{u}}$, which we represent via a joint probability density/mass function $p \left ( \mathbf{Y}^{\mathrm{obs}}, \mathbf{Y}^{\mathrm{mis}} \mid \mathbf{T}, \mathbf{Z}, \mathbf{G}^{\mathrm{o}}, \mathbf{G}^{\mathrm{u}}, \boldsymbol{\theta} \right )$ with parameter vector $\boldsymbol{\theta}$. (iv) The final input is the prior distribution for $\boldsymbol{\phi}$ and $\boldsymbol{\theta}$, denoted by $p \left ( \boldsymbol{\phi}, \boldsymbol{\theta} \right )$. We assume that $\boldsymbol{\phi}$ and $\boldsymbol{\theta}$ are distinct and do not share any parameters. We further assume the following cross-world independence condition about $\mathbf{G}$.
\begin{assumption}
    $\mathbf{G}^{\mathrm{o}}$ and $\mathbf{G}^{\mathrm{u}}$ are a priori independent of each other given $\mathbf{T}, \mathbf{Z}$. 
\end{assumption}
\noindent
This assumption eliminates the residual correlation between any two cross-world potential degree of interference functions after adjusting for the measured information, $\mathbf{T}$ and $\mathbf{Z}$. This assumption is necessary because the correlation structure between them cannot be learned from the data at hand.
We describe our priors for the DoI, potential outcomes and the parameters $( \boldsymbol{\phi}, \boldsymbol{\theta} )$ in \S\ref{sec:models}.

Given those four inputs, we first express the complete data likelihood function under Assumption \ref{asmp:unconfoundedness} and the assumption of independence in the statistical model according to
\begin{align*}
    & L \left( \boldsymbol{\theta}, \boldsymbol{\phi} \mid \mathbf{Y}^{\mathrm{obs}}, \mathbf{Y}^{\mathrm{mis}}, \mathbf{G}^{\mathrm{o}}, \mathbf{G}^{\mathrm{u}}, \mathbf{T}, \mathbf{Z}  \right) 
    \propto p \left( \mathbf{Y}^{\mathrm{obs}}, \mathbf{Y}^{\mathrm{mis}}, \mathbf{G}^{\mathrm{o}}, \mathbf{G}^{\mathrm{u}} \mid \mathbf{T},  \mathbf{Z}, \boldsymbol{\theta}, \boldsymbol{\phi} \right)\\
    &= p \left( \mathbf{Y}^{\mathrm{obs}}, \mathbf{Y}^{\mathrm{mis}}, \mathbf{G}^{\mathrm{o}}, \mathbf{G}^{\mathrm{u}} \mid \mathbf{T}, \boldsymbol{\theta}, \boldsymbol{\phi} \right) 
    % = \prod_{i=1}^N p \left( Y_{i}^{\mathrm{obs}}, \mathbf{Y}_{i}^{\mathrm{mis}}, G_{i}^{\mathrm{o}}, \mathbf{G}_{i}^{\mathrm{u}} \mid \mathbf{T}_{i}, \boldsymbol{\theta}, \boldsymbol{\phi} \right) \\
    = \prod_{i=1}^N p \left( Y_{i}^{\mathrm{obs}}, \mathbf{Y}_{i}^{\mathrm{mis}} \mid G_{i}^{\mathrm{o}}, \mathbf{G}_{i}^{\mathrm{u}},  \mathbf{T}_{i}, \boldsymbol{\theta} \right) p \left(G_{i}^{\mathrm{o}}, \mathbf{G}_{i}^{\mathrm{u}} \mid \mathbf{T}_{i}, \boldsymbol{\phi} \right).
\end{align*}
The posterior distribution of the model parameters $p\left(\boldsymbol{\phi}, \boldsymbol{\theta} \mid \mathbf{Y}^{\mathrm{obs}}, \mathbf{T}, \mathbf{Z}\right)$ is thus derived as
\[
p\left(\boldsymbol{\phi}, \boldsymbol{\theta} \mid \mathbf{Y}^{\mathrm{obs}}, \mathbf{T}, \mathbf{Z}\right)  \propto p\left(\boldsymbol{\phi}, \boldsymbol{\theta}\right) \int \prod_{i=1}^N p \left( Y_{i}^{\mathrm{obs}}, \mathbf{Y}_{i}^{\mathrm{mis}} \mid G_{i}^{\mathrm{o}}, \mathbf{G}_{i}^{\mathrm{u}}, \mathbf{T}_{i},  \boldsymbol{\theta} \right) p \left(G_{i}^{\mathrm{o}}, \mathbf{G}_{i}^{\mathrm{u}} \mid \mathbf{T}_{i}, \boldsymbol{\phi} \right) dG_{i}^{\mathrm{o}} d\mathbf{G}_{i}^{\mathrm{u}} d\mathbf{Y}_{i}^{\mathrm{mis}}. 
\]
\noindent 
% The last two expressions follow as a consequence of deFinetti's Theorem. 
The posterior distribution of the model parameters based on the complete data likelihood function is derived in a straightforward manner as
\[
p\left(\boldsymbol{\phi}, \boldsymbol{\theta} \mid \mathbf{Y}^{\mathrm{obs}}, \mathbf{Y}^{\mathrm{mis}}, \mathbf{G}^{\mathrm{o}}, \mathbf{G}^{\mathrm{u}}, \mathbf{T}, \mathbf{Z} \right)  \propto p\left(\boldsymbol{\phi}, \boldsymbol{\theta}\right) \prod_{i=1}^N p \left( Y_{i}^{\mathrm{obs}}, \mathbf{Y}_{i}^{\mathrm{mis}} \mid G_{i}^{\mathrm{o}}, \mathbf{G}_{i}^{\mathrm{u}},  T_{i}, \boldsymbol{\theta} \right) p \left(G_{i}^{\mathrm{o}}, \mathbf{G}_{i}^{\mathrm{u}} \mid T_{i}, \boldsymbol{\phi} \right). 
\]
Similarly, the posterior of the missing outcomes and DoIs conditional on the observed data and model parameters is $\prod_{i=1}^N p \left \{ \left( \mathbf{Y}_{i}^{\mathrm{mis}} \mid Y_{i}^{\mathrm{obs}}, G_{i}^{\mathrm{o}}, \mathbf{G}_{i}^{\mathrm{u}}, \mathbf{T}_{i}, \boldsymbol{\theta} \right) p \left( G_{i}^{\mathrm{o}}, \mathbf{G}_{i}^{\mathrm{u}} \mid Y_{i}^{\mathrm{obs}}, \mathbf{T}_{i}, \boldsymbol{\phi} \right) \right \}$. 

We use the Gibbs sampler to obtain posterior draws for the imputations and thereby derive the posterior distributions of the causal estimands using \eqref{eq:posterior_of_estimand} \citep{ImbensRubin1997}. Specifically, in each iteration of the sampler we alternate between drawing from the conditional posterior distributions of $\boldsymbol{\phi}, \boldsymbol{\theta}, \mathbf{G}^{\mathrm{o}}, \mathbf{G}^{\mathrm{u}}$, and $\mathbf{Y}^{\mathrm{mis}}$ given the other variables, respectively, and then use the draws of $\mathbf{Y}^{\mathrm{mis}}$ to obtain draws of the causal estimands from their respective posterior distributions. 
Details on the Gibbs sampler are in \S\ref{sec:Gibbssampling_outline}.

\begin{remark}[Formalism from a Bayesian perspective]
    Note that formalizing the DoI-based framework does not immediately imply a Bayesian perspective. We view our inferential framework as a model-assisted approach  \citep{Sarndal1992,Basse2018}, and we use our semi-parametric model as an assistant to improve precision in estimating causal effects, yet continue to rely primarily on randomization (the design) as the fundamental source of valid causal inference. We may use the frequentist tool, such as fitting a semiparametric regression of the form $Y_i = X_i \beta + G_i + \epsilon_i,$ where $\epsilon_i$ is the noise term, and $G_i$ is a flexible, nonparametric function, which can be estimated as a function of covariates and treatment status of the neighboring units, $X_{i},\X_{-i}, \A, \Z_{-i}$, using kernel regressions, spline-based approaches, local polynomial regressions, or similar methods. By flexibly capturing the interference mechanism, these methods enhance precision while design-based (or randomization-based) inference procedures preserve validity.

    %{\color{red} I would suggest significantly shortening (into three sentences) the following discussion. We do not need to promote the Bayesian paradigm.}
    
    We adopt a Bayesian perspective in this article because the Bayesian approach provides a coherent system for uncertainty quantification. Throughout this article, we also adopt a finite-population perspective for causal inference, where $Y_i(\mathbf{z})$ is treated as a fixed, non-random quantity. From the Bayesian perspective, the ``model'' is not merely about sampling from a super-population; rather, it structures our prior beliefs about outcomes in the finite population, enabling us to incorporate uncertainty about the interference mechanism (see discussions in Section \ref{sec:structural_assumption}). 
\end{remark}

\subsection{Bayesian Semiparametric Models}
\label{sec:models}

The models in our Bayesian approach consist of a combination of parametric and nonparametric specifications. We adopt a semi-parametric model for the $Y$-model, with the particular form specified based on the specific setting under consideration. We shall discuss the specification of the $Y$-model later in our simulation studies and case study. In those studies, we set $\boldsymbol{\theta} = (\beta, \lambda)^\top$, where $\beta$ contain regression coefficients associated with the covariates and $\lambda$ is a scale parameter in the regression. The prior for $(\beta,\lambda)$ is the Normal-Inverse-Gamma distributions, $\mathrm{N}(0,\sigma_1^2)\mathrm{IG}(a_1,b_1)$ with $\sigma_1^2$, $a_1$ and $b_1$ being the variance, shape and scale parameters respectively. 
% Throughout the manuscript, we will adopt the same parameterizations for the corresponding distributions. 

The interference structure captured by the DoI is generally complex, hence unlikely to be adequately represented by parametric models. 
% Furthermore, as the DoI are latent variables, if we were to adopt parametric models for them it would be difficult to diagnose their validity or modify them if misspecification of the $G$-model is a concern. 
As such, we utilize a Bayesian nonparametric prior for the $G$-model, specifically, the Dependent Dirichlet process mixture prior (DDPM). As indicated by equation \eqref{eq:eff_tr_g}, the estimation of the causal estimands is reduced to estimation of the probability density function of the DoI, and this further justifies our use of the DDPM as it is a natural choice for density estimation under the Bayesian paradigm. In addition to its flexibility, the clustering property of the DDPM, in terms of its potential ability to automatically separate units into several clusters based on how they interfere with one another, provides another advantage that motivates our adoption of this nonparametric model. \citet{Fernando2022} provides a comprehensive review of the DDPM. 

A random probability measure $H$ drawn from a Dirichlet Process, $\mathrm{DP}(\alpha, H_0)$, with concentration parameter $\alpha>0$ and base probability measure $H_0$ over a measurable space $(\Phi, \mathcal{B})$  is such that for any finite partition $(B_1, \ldots ,B_k)$ of $\mathcal{B}$, $\left ( H(B_1), \ldots, H(B_k) \right ) \sim \mathrm{Dir} \left ( \alpha H_0(B_1), \ldots, \alpha H_0(B_k) \right )$, where $\mathrm{Dir}(\alpha_1,...,\alpha_k)$ denotes the Dirichlet distribution with parameters $\alpha_1,...,\alpha_k > 0$ \citep{Ferguson1974}. The concentration parameter $\alpha$ determines the number of clusters in the Dirichlet Process.
Our DDPM prior for the probability density function of the $G_i$ is specified as 
$$(G_1, \ldots, G_N) \mid H_{\mathbf{T}, \mathbf{Z}} \displaystyle{\indsim} p(G \mid H_{\mathbf{T}, \mathbf{Z}}),$$ where $p(G \mid H_{\mathbf{T}, \mathbf{Z}}) = \int \eta(G \mid \phi) dH_{\mathbf{T}, \mathbf{Z}}(\phi)$ with $\eta(\cdot \mid \phi)$ being a continuous density function for every $\phi \in \Phi$, and $H_{\mathbf{T}, \mathbf{Z}} \sim \mathrm{DP}(\alpha, H_0)$ for some $\alpha > 0$ and a base measure $H_0$ \citep{Maceachern2000}, where $\displaystyle{\indsim}$ indicates independently distributed variables.
The subscript of $H_{\mathbf{T}, \mathbf{Z}}$ indicates the dependence of the measure on $\mathbf{T}$ and $\mathbf{Z}$. As shown below, we include covariate information in the DDPM by allowing the locations in the stick-breaking construction of the $\mathrm{DP}(\alpha, H_0)$ to depend on covariates. We consider using common weights across the values of DoI, so that our DDPM is a ``single-weights'' DDPM, for its simplicity and ease of implementation (which follows as it ignores the dependence of the weights on covariates). The representation of the stick-breaking process is $H_{\mathbf{T}, \mathbf{Z}} = \sum_{k=1}^{\infty} w_k \delta_{\phi_k (\mathbf{T}, \mathbf{Z})}(\cdot) $, $w_k = v_k \prod_{l<k}(1-v_l)$, and $v_l \indsim \mathrm{Beta}(1, \alpha)$. Ultimately, our $G$-model is
\begin{equation}
\label{eq:dpm}
p \left ( G_i \mid \mathbf{T}_i, \Z_{-i}, \boldsymbol{\phi} \right ) \propto \sum_{k=1}^{\infty} w_k \left ( \sigma_k^2 \right )^{-1/2} \mathrm{exp} \left ( -\frac{1}{2\sigma_k^2} \left ( G_i - \mu(\mathbf{T}_i, \Z_{-i}, \gamma_k) \right )^2 \right )
\end{equation}
where the atoms $\boldsymbol{\phi}_k = \left ( \gamma_k, \sigma_k^2 \right )$  and the weights $w_k$ are nonparametrically specified via the $\mathrm{DP}(\alpha, H_0)$. This is an infinite mixture of Normal distributions with $\mu( \mathbf{T}_i, \Z_{-i}, \gamma_k)$ being the location parameter of each component that is a function of the covariates of unit $i$, the $i$th row of $A$, and treatment assignments excluding that of unit $i$. 

Finally, we specify the priors for $\alpha$ and $H_0$. We use the $\mathrm{Gamma}(a,b)$ prior for $\alpha$ with shape $a$ and scale $b$ \citep{Escobar1995}, and obtain posterior draws of $\alpha$ via a Metropolis-Hastings step in our Gibbs sampler. Next,  we take $H_0$ as a Normal-Inverse-Gamma conjugate $\mathrm{N}(\mu_0,\sigma_0^2)\mathrm{IG}(a_0,b_0)$ with $a_0$ and $b_0$ being the shape and scale parameter respectively. Our simulation studies and case studies illustrate how the values for all the hyperparameters are set.

\begin{remark}[The role of $\mu$ with domain knowledge about interference]
    The DDPM offers a straightforward way to model distributions that evolve depending on covariates. In our case, using the DDPM to represent the uncertainty of the interference structure (as detailed in \S\ref{sec:DOI}) allows this uncertainty to evolve based on characteristics such as the network matrix, treatment variables, and covariates, which are typically extremely high-dimensional. 
    Therefore, following \citet{Maceachern1999}, we advocate for specifying the dependence on the uncertainty distribution by lower-dimensional summaries of these characteristics, incorporating the analysts' domain knowledge about interference. In this context, we specify $\mu$ to incorporate our possibly limited understanding of the interference structure.
    By specifying $\mu$, researchers can leverage their domain expertise to develop network features that are expected to influence interference. 
    On the other hand, while specifying $\mu_k$ is necessary, the precise choice of $\mu_k$ is not as critical. This is because the resulting marginal distribution for $G_i$ forms a flexible, infinite mixture distribution, as detailed in Equation \eqref{eq:dpm}. The key advantage of adopting the DDPM for the $G$-model lies in this nonparametric nature. In our simulation studies in \S\ref{sec:simulation_studies}, we have verified that even basic forms of $\mu_k$ are effective in handling complex interference structures.
\end{remark}

\subsection{Structural Assumptions for Inference and Prior Elicitation}
\label{sec:structural_assumption}
The DoI framework is conceptually straightforward, gives us a clear interpretation of the treatment effects under interference, and extends conventional exposure maps by allowing itself to be unknown, as discussed in \S\ref{sec:DOI}. However, this flexibility comes with a trade-off, necessitating an additional structural assumption between $\mathbf{Y}$ and $\mathbf{G}$. Specifically, when $G_i$ is fully latent, Assumption \ref{asmp:doi} remains non-restrictive because any function mapping each value of the treatment assignment vector to a unique real number will satisfy the assumption. This is true because the cardinality of the support of the treatment assignment vector is always less than or equal to that of the real numbers. As \S\ref{sec:methodology_overview} implies, statistical inference regarding the causal effects requires some modeling assumptions between $\mathbf{Y}$ and $\mathbf{G}$.

This structural assumption can be tailored to specific applications, allowing the integration of prior knowledge into the inference model. For example, one can view each unit’s outcome $Y_i$ as determined by its own treatment $Z_i$ and a latent function $G_i$ that collects how other units’ treatments, covariates, and the network structure affect $Y_i$. Formally, $Y_i \;=\; f_i\left(Z_i,\;G_i(X_i, \X_{-i}, \A, \Z_{-i}),\;\epsilon_i\right),$ where $\epsilon_i$ is an exogenous noise term and $G_i$ is nonparametric, flexibly characterizing the interference channel from other units. Assumption \ref{asmp:doi} (equating potential outcomes whenever $G_i$ coincides in distribution) then says that if two configurations of $\{X_i, \X_{-i}, \A, \Z_{-i}\}$ yield the same effective distribution for $G_i$, they are indistinguishable to $Y_i$.            
From a more classical statistical modeling viewpoint, the exposure map may be interpreted as a \emph{fixed effect} (i.e., deterministic function of the observed data), while the DoI can be viewed as a \emph{random effect} that encapsulates additional, possibly unobserved, sources of variation. Consequently, if we specialize $f_i$ to be a linear function in its arguments, then the DoI framework reduces to a linear mixed model, with $G_i$ playing the role of a random component that depends on the network structure and other units’ treatments.

% For instance, we may consider an additive structure for the main and spillover effects:
% $Y_i(z,\mathbf{z}_{-i}) = f_z(X_i) + g(\mathbf{T}_i, \mathbf{z}_{-i}) + \epsilon_i$,
% where $f_z$ is a real-valued function for $Z_i=z$ that defines the direct effect, $g$ is a real-valued function representing the degree of interference for unit $i$ that defines the spillover effect, and $\epsilon_i$ is an appropriate error term. 
% \yo{Considering the relationship between the DoI framework and the exposure mapping approach from different perspectives, a parallel can be drawn: the exposure map can be interpreted as a fixed effect, while the DoI represents a flexible random effect specified nonparametrically. The nonparametric nature of the random effect introduces added flexibility in inference by allowing dependencies on covariates and the network structure.}

A key advantage of our DoI framework, with its structural assumption, over traditional exposure mapping is the clear separation of roles. As \citet{Fredrik2021_2} notes, exposure mapping in existing literature serves a dual purpose: defining direct and spillover effects and imposing assumptions on the interference structure. This dual role necessitates defining both the interference structure, which determines a unique potential outcome, and exposure effects, which determine causal quantities. Given that exposure mapping is problem-specific, its effects often lack scientific meaning outside the given problem. Moreover, misspecification of these maps typically leads to biased estimators \citep{Fredrik2021_2}.

This separation enables a unified procedure in our inferential approach, allowing for the incorporation of case-specific interference structures in estimating causal effects without altering the effects being estimated. This is because our causal estimands are generally defined in terms of treatment assignments and are not dependent on exposures. 
Secondly, we incorporate knowledge about interference structures by specifying priors on the DoI, capturing their complexities and uncertainties, and yielding valid posterior distributions for finite-population causal estimands. The Bayesian methodology propagates uncertainties about interference structures using an imputation-based approach, as outlined in \S\ref{sec:methodology_overview}. Our approach distinguishes the interference structure from the effects of interest, as our causal estimands are generally defined in terms of treatment assignments and are not dependent on exposures.

To motivate the idea of prior elicitation given prior investigation or knowledge about the interference structure, we discuss how to incorporate this knowledge in the form of the DoI. Suppose we know that the interference structure satisfies the stratified interference condition \citep{Hudgens2008}, which states that spillover effects are determined solely by the proportion of treated units within the same cluster as each unit. In this case, we may specify the function $\mu(\mathbf{T}_i, \Z_{-i}, \gamma)$ for unit $i$ as  
\[
\mu(\mathbf{T}_i, \Z_{-i}, \gamma_k) = \gamma_k \frac{\sum_{j \in \mathcal{G}_i} Z_j}{|\mathcal{G}_i|},
\]  
where $\mathcal{G}_i$ is the connected subgroup to which unit $i$ belongs, and $|\mathcal{G}_i|$ is the size of $\mathcal{G}_i$. If we have more detailed knowledge that, for all $i$, unit $i$ is affected by its direct neighbors and $m$-th order neighbors in the same connected graph, but is affected by them differently, we may specify  
\[
\mu(\mathbf{T}_i, \Z_{-i}, \gamma_k) = \sum_{s=1}^{m}\gamma_{s,k} \frac{\sum_{j \in \mathcal{G}_i^{(s)}} Z_j}{|\mathcal{G}_i^{(s)}|},
\]  
where $\mathcal{G}_i^{(s)}$ denotes the set of $s$-th order neighbors (neighbors at distance $s$) of unit $i$, and $|\mathcal{G}_i^{(s)}|$ is its corresponding size. If we wish to include additional demographic information about the neighbors, we may adjust the function accordingly (e.g., the number of teenagers, etc.).  

If individuals have even deeper insights into the interference structure and know that spillover effects are determined by certain metrics of the neighbors that represent the local importance of units (e.g., centrality, eigenvalues, PageRank metrics \citep{Newman2010}, etc.), as is often the case in social network analysis, we may incorporate these metrics into the specification of $\mu(\mathbf{T}_i, \Z_{-i}, \gamma_k)$. For example,  
\[
\mu(\mathbf{T}_i, \Z_{-i}, \gamma_k) = \gamma_{1,k} \sum_{j \neq i} P_j \exp(-\gamma_{2,k}d(i,j)),
\]  
where $P_j$ represents the importance metric of unit $j$, and $d(i,j)$ represents the (potentially weighted) distance between units $i$ and $j$ in the network. The term $\exp(-\gamma_{2,k}d(i,j))$ acts as a decaying factor for the influence of unit $j$ on unit $i$.  
% These flexible specifications of the interference structure are not permitted within the exposure mapping framework because the resulting exposure effects are not straightforward to interpret. 
% This limitation arises because the exposure mapping framework defines the interference structure and estimands simultaneously, restricting its flexibility. 
% {\color{red} I'd remove the last two sentences.}

\begin{example}[Linear-in-Means Model]
The Linear-in-Means (LIM) model is a social interaction model that is commonly used in the economic literature \citep{BRAMOULLE200941,Chin2019,Bramoulle2020,Leung2021}. Assuming $|\gamma| < 1$, the model defines potential outcomes $Y_i(\mathbf{Z})$ through
\begin{equation*}
    \mathbf{Y} = \alpha + \beta \mathbf{Z} + \gamma\Tilde{\mathbf{A}}\mathbf{Y} + \delta \Tilde{\mathbf{A}} \mathbf{Z} + \boldsymbol{\epsilon},
\end{equation*}
where $\boldsymbol{\epsilon}=(\epsilon_1, \ldots, \epsilon_N)^\top$ and $\Tilde{\mathbf{A}}$ is obtained by row-normalizing $\mathbf{A}$ (divide each row by its sum). By the matrix identity $(\mathbf{I}-\gamma \Tilde{\mathbf{A}})^{-1} = \sum_{k=0}^{\infty}\gamma^{k}\Tilde{\mathbf{A}}^{k}$, one obtains the unit-level reduced form of the model \citep{BRAMOULLE200941}: 
\begin{equation}
\label{eq:LIM_model}
    Y_i = \Tilde{\alpha} + \Tilde{\beta} Z_i +  \sum_{k=0}^{\infty}\Tilde{\gamma}_k (\Tilde{\mathbf{A}}^{k+1} \mathbf{Z})_i + \Tilde{\epsilon}_i,   
\end{equation}
where  $\Tilde{\alpha}=\alpha/(1-\gamma)$, $\Tilde{\beta} = \beta$, $\Tilde{\gamma}_k = (\beta\gamma + \delta)\gamma^k$, $\Tilde{\epsilon} = \sum_{k=0}^{\infty}\gamma^k \mathbf{A}^{k} \boldsymbol{\epsilon}$, and $(\Tilde{\mathbf{A}}^{k+1} \mathbf{Z})_i$ is the $i$-th cordinate of $\Tilde{\mathbf{A}}^{k+1} \mathbf{Z}$. The third term indicates that the impact of treatments assigned to $k$-neighbors is exponentially down-weighted by $\Tilde{\gamma}_k$. Our additive semi-parametric model specification is a general form of the LIM model because \eqref{eq:LIM_model} is expressed as a linear combination of the direct effects ($\Tilde{\beta} Z_i$ and the term that corresponds to $Z_i$ in $\sum_{k=0}^{\infty}\Tilde{\gamma}_k (\Tilde{\mathbf{A}}^{k+1} \mathbf{Z})_i$) and the infinite sum of the spillover effects (the terms that correspond to $\mathbf{Z}_{-i}$ in $\sum_{k=0}^{\infty}\Tilde{\gamma}_k (\Tilde{\mathbf{A}}^{k+1} \mathbf{Z})_i$), and the infinite sum of the spillover can correctly be specified by the infinite mixture of the $G$-model \eqref{eq:dpm}.

\end{example}

\begin{remark}[Extension to more general interaction structures]
Although we focus here on network-based adjacency information (which typically encodes pairwise edges between units), the DoI framework is not limited to strictly pairwise (graphical) interactions. In principle, one can generalize the set $\mathcal{A}$ to represent hypergraphs or other higher-order structures in which ``edges'' (hyperedges) connect subsets of more than two units.

Concretely, suppose we have a hypergraph $\mathcal{H} = \{ e_1, e_2, \dots, e_K \}$, where each hyperedge $e_k \subseteq \{1,2,\dots,N\}$ is a subset of units. The usual adjacency matrix $\mathbf{A}$ in a graph may be replaced by an incidence structure for hypergraphs, for instance, an incidence matrix $\mathbf{B} \in \{0,1\}^{N \times K}$ where
\[
    B_{i,k} = 
    \begin{cases}
    1, & \text{if unit } i \in e_k, \\
    0, & \text{otherwise}.
    \end{cases}
\]
Incorporating such a higher-order structure into the DoI framework means that each unit $i$ has access to information not just about pairwise connections with others but also about membership in shared hyperedges. Thus, the characteristic set $\mathcal{T}$ can be expanded to capture hypergraph membership. Formally, we can write $\mathcal{T} = \mathcal{X} \times \mathcal{X}^{N-1} \times \mathcal{H},$
where $\mathcal{H}$ denotes the set of all possible hypergraph incidence structures (analogous to how $\mathcal{A}$ previously denoted the set of all adjacency matrices).

Then, the DoI for unit $i$ can be modified to $G_i: \left(\mathcal{X} \times \mathcal{X}^{N-1} \times \mathcal{H}\right) \times \Omega^{N-1} \to \mathbb{R},$
so that when evaluating $G_i$, one accounts for all relevant hyperedges $e_k$ that involve unit $i$ (or subsets thereof that might exert a joint influence on $i$). For example, if one models a situation where interference depends on whether multiple distinct units are jointly exposed to certain treatments, these higher-order relationships can be captured by the hyperedges in $\mathcal{H}$. For instance, one may specify a mean function
\begin{align*}
    \mu(\mathbf{T}_i, \mathbf{Z}_{-i}, \gamma_j) =  \sum_{k=1}^{K} \gamma_{j,k} \mathbf{1}\{i \in e_k\} \frac{\sum_{j \in e_k,\, j \neq i} Z_j}{|e_k|},
\end{align*}
which measures the proportion of other treated units in each hyperedge $e_k$. 
This flexibility allows the DoI framework to model a richer class of interference structures than what can be described by a standard adjacency matrix, accommodating any scenario in which the influence on a given unit depends on group-level attributes or actions of other units, rather than strictly pairwise interactions.
\end{remark}

\subsection{Details on the Gibbs Sampler}
\label{sec:Gibbssampling_outline}

A discussion on efficient MCMC algorithms for DDPM models is provided by \citet{Escobar1994}. Our $Y$-model is generally specified by $\boldsymbol{\theta}$ as $p(Y_i(z,g_i) \mid \mathbf{T}_{i}, G_i=g_i, \boldsymbol{\theta})$. For our DDPM, we utilize an approximated blocked Gibbs sampler based on a truncation of the stick-breaking representation of the DP proposed by \citet{Ishwaran2000}. This algorithm proceeds by first selecting a conservative upper bound on the number $K$ of clusters. Let $C_i \in \{1,...,K\}$ denote the latent class indicators for unit $i$. We specify a Multinomial distribution $C_i \sim \mathrm{MN}(\mathbf{w})$ on $C_i$,  where $\mathbf{w} = (w_1,...,w_K)^\top$ contains the weights from the DDPM. Conditional on $C_i=k$, equation \eqref{eq:dpm} is simplified to a single Normal component. \citet{Ishwaran2001} demonstrated that an accurate approximation to the exact DP is obtained as long as $K$ is sufficiently large. To ensure this, we ran several MCMC iterations with different values of $K$ and increased $K$ after an iteration if all clusters were occupied. We terminated this process when the number of occupied clusters was less than $K$. 
Our entire algorithm proceeds as follows; throughout, `$\mid - $' indicates conditional on all others.
\begin{enumerate}
    \item Given $\boldsymbol{\theta}$, $\boldsymbol{\phi}$ and $\mathbf{C}=(C_1,...,C_N)$, draw each $G_i^o$ from 
    \begin{equation*}
        p\left(G_i^o \mid - \right) \propto p\left(Y_i^{\mathrm{obs}} \mid  \mathbf{T}_{i}, G_i^o, \boldsymbol{\theta} \right) p\left(G_i^o \mid \mathbf{T}_{i},  \Z_{-i}, \phi_{C_i}\right).
    \end{equation*}
    \item Given $\boldsymbol{\phi}$, $\mathbf{w}$ and $G_i^o$, draw each $C_i$ from $p\left(C_i=k \mid - \right) \propto w_k p\left(G_i^o \mid \mathbf{T}_{i},  \Z_{-i}, \boldsymbol{\phi}_k \right).$
    \item Let $w_K'=1$. Given $\alpha$, $\mathbf{C}$, draw $w_k'$ for $k \in \{1,...,K-1\}$ from
    $w_k' \sim \text{Beta} \big(1+ \sum_{i:C_i=k}1, \alpha+ \sum_{i:C_i>k}1 \big)$.
    Then, update $w_k= w_k'\prod_{j<k}(1-w_j')$ for $k=1,\ldots, K$.
    \item Given $\mathbf{C}$ and $\mathbf{w}'$, draw $\alpha$ from $p \left(\alpha \mid - \right) \propto p \left( \alpha \right) \prod_{k=1}^{K} f \left(w_k' \mid 1+ \sum_{i:C_i=k}1, \alpha+ \sum_{i:C_i>k}1 \right)$
    where $f$ is the pdf of $w_k'$, the beta distribution. 
    \item Given $\mathbf{C}$ and $\mathbf{G}^o$, draw $\boldsymbol{\phi}_k$ from $p\left(\boldsymbol{\phi}_k \mid -\right) \propto H_0(\boldsymbol{\phi}_k) \prod_{i:C_i=k} p\left(G_i^o \mid \mathbf{T}_{i},  \Z_{-i}, \boldsymbol{\phi}_k \right).$
    \item Given $\boldsymbol{\phi}$, $\mathbf{C}$ and $\mathbf{G}^o$, draw $\boldsymbol{\theta}$ from $p \left(\boldsymbol{\theta} \mid - \right) \propto p(\boldsymbol{\theta}) \prod_{i=1}^{N} p\left(Y_i^{\mathrm{obs}} \mid  \mathbf{T}_{i}, G_i^o, \boldsymbol{\theta} \right).$
\end{enumerate}
\noindent  The imputations of the $\mathbf{Y}_i^{\mathrm{mis}}$ are generated from the $Y$-model using the $G_i^o$ in Step 1 and the $\boldsymbol{\theta}$ in Step 6. Specifically, for $i=1, \ldots ,N$:
\begin{enumerate}\addtocounter{enumi}{6}
    \item Given $G_i^o$ in Step 1 and $\boldsymbol{\theta}$ in Step 6, draw from $Y_i(z', \z_{-i}) \sim p\left(Y_i \mid Z_i=z', \mathbf{T}_{i}, G_i^o, \boldsymbol{\theta} \right).$
\end{enumerate}
To obtain posterior draws of the spillover effect estimands, we impute the $Y_i(z, \z_{-i}')$ by using the posterior draws of $G_i^u=G(\mathbf{T}_i, \z')$ and $\boldsymbol{\theta}$. 
\begin{enumerate}\addtocounter{enumi}{7}
    \item Given $\mathbf{w}$ in step 3, draw $C_i^{\mathrm{u}}$ from $C_i^{\mathrm{u}} \sim \text{MN}(\mathbf{w}).$
    \item Given $C_i^{\mathrm{u}}$ and $\boldsymbol{\phi}$, draw $G_i^{\mathrm{u}}$ from $G_i^{\mathrm{u}} \sim \text{N}(\mu(\mathbf{T}_i, \z_{-i}', \gamma_{C_i^{\mathrm{u}}}),\sigma_{C_i^{\mathrm{u}}}^2).$
    \item Given $G_i^{\mathrm{u}}$ in step 9 and $\boldsymbol{\theta}$ in step 6, draw from $Y_i(z, \z_{-i}') \sim p\left(Y_i \mid Z_i=z, \mathbf{T}_{i}, G_i^u, \boldsymbol{\theta} \right).$
\end{enumerate}
This Gibbs sampler ultimately enables us to impute all of the missing potential outcomes that define the assignment-conditional effects, and yields approximations for the expected effects. Additional technical details regarding these points and each step are provided in the Appendix.
% For $\tau_{EASE}$, $\z_{-i}^{\mathrm{mis}}$ is set to $\zero_{N-1}$ by definition. 

% \subsection{Importance Resampling}
% TODO

\section{Simulation Studies}
\label{sec:simulation_studies}

We perform simulation studies to evaluate the performance of our Bayesian methodology under our DoI framework with respect to that of the Horvitz-Thompson (HT) estimator in the case of a Bernoulli trial.
% , in which each experimental unit is assigned to one of two levels of the treatment factor with a fixed probability. 
The HT estimator was selected as the comparison method in our evaluation because \citet{Fredrik2021} and \citet{Shuangning2022} have demonstrated that it is a consistent estimator for the E-ATE under a Bernoulli trial under interference and that it possesses desirable large sample properties.

To compare these methods, we assess the bias, mean square error (MSE), and coverage of a procedure based on $M$ simulated datasets by calculating $\sum_{n=1}^{N_{\mathrm{sim}}} \left ( \tau - \hat{\tau}_n \right )/N_{\mathrm{sim}}$, $\sum_{n=1}^{N_{\mathrm{sim}}} \left ( \tau - \hat{\tau}_n \right )^2/N_{\mathrm{sim}}$ and $\sum_{n=1}^{N_{\mathrm{sim}}} \mathbbm{1} \left ( \hat{\tau}_n^{l} \leq \tau \leq \hat{\tau}_n^{u} \right )/N_{\mathrm{sim}}$ respectively, 
where $\tau$ denotes the true causal estimand,  and $\hat{\tau}_n$, $\hat{\tau}_n^{l}$ and $\hat{\tau}_n^{u}$ are the estimates of the causal estimand, $2.5\%$ lower endpoint of the interval estimator, and $95\%$ upper endpoint of the interval estimator in dataset $n = 1, \ldots, N_{\mathrm{sim}}$. 
Here, we take the posterior mean of the causal estimand as our Bayesian point estimator and the $95\%$ central credible interval as the interval estimator. For the HT estimator under the Bernoulli trial, \citet{Fredrik2021} gave the variance estimator and conservative confidence interval. As the true value of E-ATE is hard to obtain in closed form, we approximated it in our simulation study via the Monte Carlo average of $1000$ draws for each generated dataset and graph.

We consider seven different data-generating mechanisms and two different underlying networks of experimental units. The seven data-generating mechanisms that we consider are:
\begin{enumerate}
\item $Y_i = \mathbf{X}_i^\top\beta_1 + Z_i\tau + \psi_1\sum_{j \in \mathcal{N}_i} Z_jA_{ij}/\left ( |\mathcal{N}_i|+1 \right ) + \epsilon_i$,
\item $Y_i =  \mathbf{X}_i^\top\beta_1 + Z_i\tau +  \psi_1\sum_{j \in \mathcal{N}_i} P_jZ_jA_{ij} / \left ( |\mathcal{N}_i|+1 \right ) + \epsilon_i$,
\item $Y_i = \left (\mathbf{X}_i^\top\beta_1 + \psi_1\sum_{j \in \mathcal{N}_i} P_jZ_jA_{ij} / \left ( |\mathcal{N}_i|+1 \right ) \right) \exp \left ( -\psi_2 Z_i \sum_{j \in \mathcal{N}_i} P_jZ_jA_{ij} / \left (|\mathcal{N}_i|+1 \right ) \right ) + \epsilon_i$,
\item $Y_i = \left ( \mathbf{X}_i^\top\beta_1 + Z_i\tau + \psi_1\sum_{j \in \mathcal{N}_i} P_jZ_jA_{ij} / \left ( |\mathcal{N}_i|+1 \right ) \right ) \exp \left ( -\psi_2 \sum_{j \in \mathcal{N}_i} P_jZ_jA_{ij} / \left (|\mathcal{N}_i|+1 \right ) \right ) + \epsilon_i$,
\item $Y_i = \left ( \mathbf{X}_i^\top\beta_1 + Z_i\tau + \psi_1\sum_{j \in \mathcal{N}_i} P_jZ_jA_{ij} / \left ( |\mathcal{N}_i|+1 \right ) \right ) \cos \left ( \pi \psi_2 \sum_{j \in \mathcal{N}_i} P_jZ_jA_{ij} / \left (|\mathcal{N}_i|+1 \right ) \right ) + \epsilon_i$,
\item $Y_i = \mathbf{X}_i^\top\beta_1 + \psi_2 U_i + Z_i\tau U_i + \psi_1 U_i \sum_{j \in \mathcal{N}_i} Z_jA_{ij}/\left ( |\mathcal{N}_i|+1 \right ) + \epsilon_i$,
\item $Y_i = \mathbf{C}_i^\top\beta_2 + Z_i\tau + \psi_1\sum_{j \in \mathcal{N}_i} Z_jA_{ij}/\left ( |\mathcal{N}_i|+1 \right ) + \epsilon_i$,
\end{enumerate}
where $\beta_1=(-1,1.5)^\top$, $\beta_2=(-1,1.5,0.5,-0.5,0.6,-0.6,0.8,-0.8,1.0,-1.0)^\top$, $\tau=3$, $\psi_1=2$, $\psi_2=0.5$, $\mathbf{X}_i \in \mathbb{R}^2$ has first entry $X_{i,1} = 1$ and second entry $X_{i,2} \sim \mathrm{N} \left ( 0, 1 \right )$, $\mathbf{C}_i \in \mathbb{R}^{10}$ has first entry $C_{i,1} = 1$ and other entries $C_{i,j} \sim \mathrm{N} \left ( 0, 1 \right )$ for $j=2,\ldots,10$, $Z_i \sim \text{Bernoulli}(0.5)$, $\epsilon_i \sim \text{N}(0,1)$ is an additive error term, $A_{ij}$ is the $(i,j)$ entry of the adjacency matrix, $\mathcal{N}_i$ is the set of neighboring units for unit $i$, $P_i$ denotes the eigenvector centrality measure \citep[p.~169]{Newman2010}, and  $U_i \sim \text{N}(1.0,0.5)$ represents an unmeasured covariate.  The $P_i$ is considered as a proxy for the ``importance'' of unit $i$ on the network. We assume that this measure quantifies the interference structure and spillover effects for each unit. This metric evaluates the importance of a node in a graph, based not only on the number of its direct connections (degree) but also on the centrality of its connected nodes. By accounting for these indirect influences, it effectively captures the complex spillover effects common in real-world applications, such as social network analysis, where interference issues frequently arise.
Scenario 1 corresponds to the stratified interference structure. Including $P_i$ in Scenarios 2--5 creates more complicated interference structures because relative scores are assigned to all nodes within the network, based on the principle that connections to high-scoring nodes contribute more significantly to a node's score than identical connections to low-scoring nodes. Untreated units are down-weighted by $0.1$ when computing $P_i$. Furthermore, the treatment effect in Scenario 3 has a nonlinear multiplicative form in that the treatment variable $Z_i$ appears in the exponential function. Additionally, Scenarios 4 and 5 have additive and multiplicative treatment and spillover effects. Scenario 6 examines the interference pattern involving an unmeasured variable $U_i$, which is not used for inference. In this scenario, the outcome $Y_i$ depends on $U_i$ both directly and through its role in modifying the effects of treatment and spillover. It is important to note that $U_i$ is not necessarily an unmeasured confounder, as the treatment assignment does not depend on $U_i$. Finally, Scenario 7 examines the performance of the proposed method for a moderate-sized dimension of covariates.

To see the performance of the methodologies under different underlying networks, we use two different random graph models to generate the underlying networks: the Erd\"os-R\'enyi and Barab\'asi–Albert graphs \citep{frieze_karonski_2015}.
In all seven scenarios, we simulate 
$N_{\mathrm{sim}} = 200$ datasets for each sample size of $N = 300, 500, 1000, 2000, 3000$.  
We ran the MCMC algorithm for $2000$ iterations after a burn-in of $2000$. 
We chose the iteration numbers after experimentation verifying they deliver stable results over multiple runs. 

For scenario 1--6, the $Y$-model is specified by $\boldsymbol{\theta} = (\beta, \lambda)^\top$ as: $Y_i(z) \sim \mathrm{N}(\mathbf{X}_i^\top \boldsymbol{\beta}_z +  G_i(\mathbf{T}_i, \mathbf{Z}_{-i}), \lambda_{z})$. In the $G$-model we set $\mu(\mathbf{T}_i, \Z_{-i}, \gamma_k) = \gamma_{k,1} + \gamma_{k,2} \sum_{j \in \mathcal{N}_i} Z_jA_{ij}$. The prior distributions for the parameters in this case are $\boldsymbol{\beta}_{z} \sim \mathrm{N}((0,0)^\top, 10^2\mathbf{I}_2)$, where $\mathbf{I}_2$ is the $2 \times 2$ identity matrix, $\lambda_z \sim \text{IG}(0.1,0.1)$, $\gamma_{k,1}, \gamma_{k,2} \sim \text{N}(0,10^2)$ and $\sigma_k^2 \sim \text{IG}(0.1,0.1)$. For scenario 1, the inference model is well specified for the additive interference structure of the $Y$-model as well as the $G$-model in \S\ref{sec:structural_assumption}.  
For scenarios 2, the $Y$-model has a correct additive form but the location function $\mu_k$ of the $G$-model is simplified compared to the true data-generating process. Specifically, it considers only the treatment status of the neighboring units and ignores the spillover effects from $k$-neighbors for $k \geq 2$ in its specification. This scenario assesses the robustness of our methodologies for a specific, simple choice of $\mu_k$.
For scenarios 3--5, the inference model is misspecified in that the data-generating models are multiplicative of treatment and spillover effect terms ($\exp$ and $\cos$), whereas the inferential model is additive. These four scenarios assess the robustness of our approach to structural misspecification. Scenario 6 further evaluates the robustness of the proposed methodology in the presence of an unmeasured covariate. For Scenario 7, the lower-dimensional covariates $X_i$ are replaced with higher-dimensional covariates $C_i$ and the dimension of $\boldsymbol{\beta}_{z}$ increases accordingly.

\begin{figure*}
\centering
\includegraphics[scale=0.6]{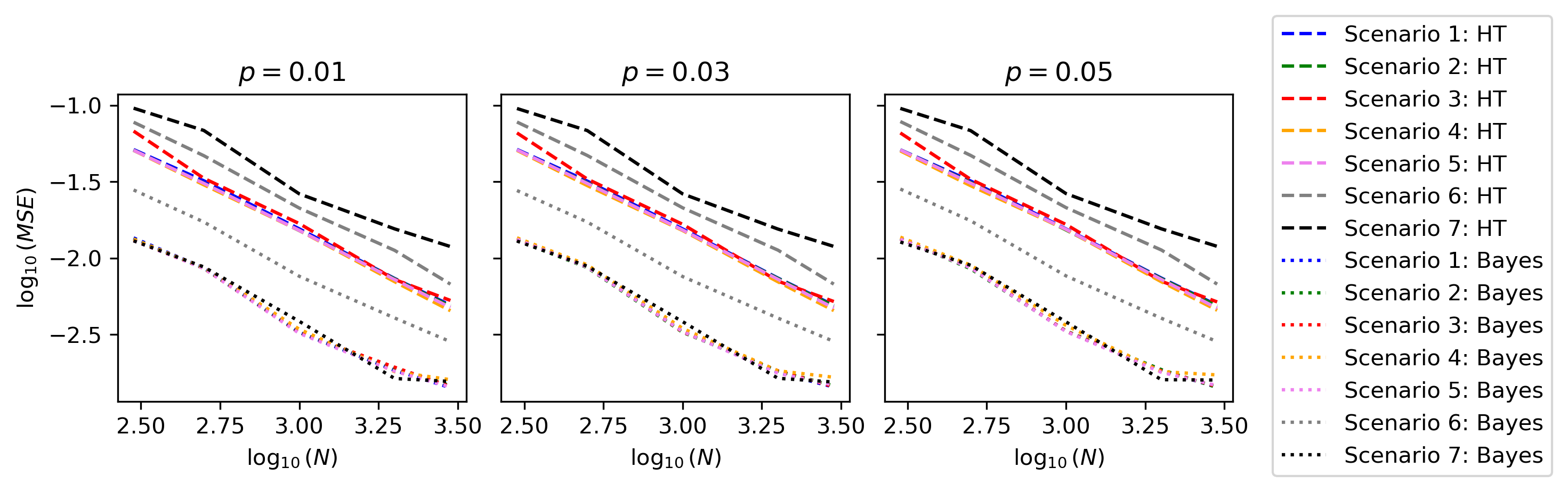}
\caption{MSE of the HT estimator and our proposed Bayesian methodology for the Erd\"os-R\'enyi graphs $\mathrm{ER}(N,p)$ with different sparsity parameters $p$.}
\label{fig:MSE_ER}
\end{figure*}

\begin{figure}
\centering
\includegraphics[scale=0.6]{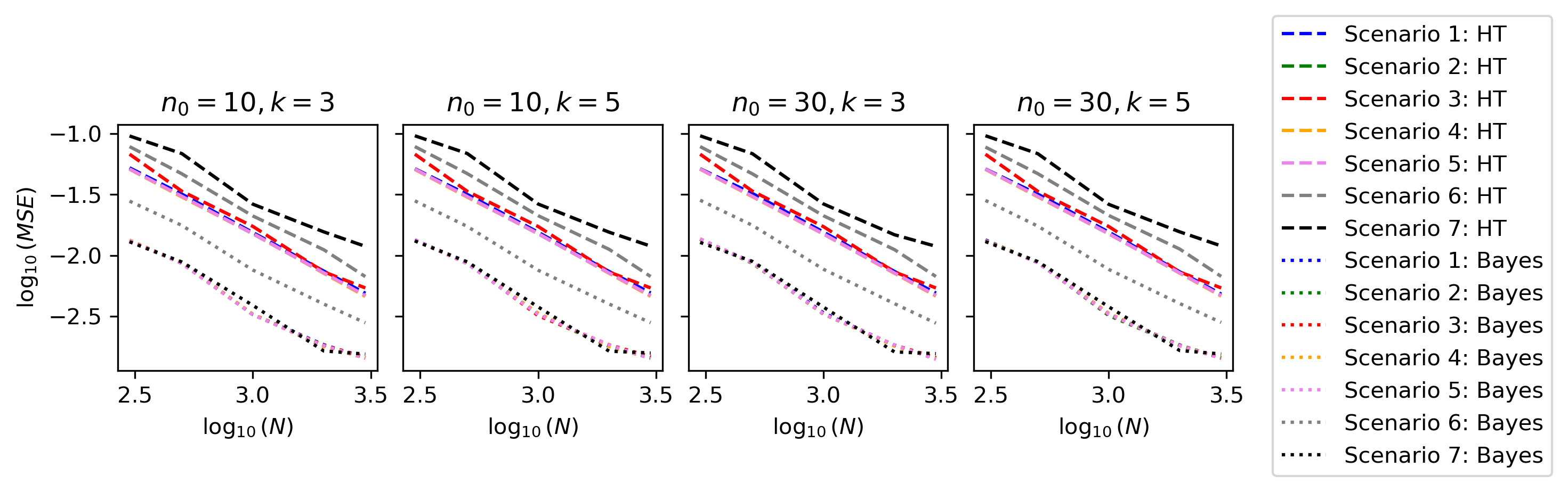}
\caption{MSE of the HT estimator and our proposed Bayesian methodology for the Barab\'asi–Albert graphs $\mathrm{BA}(N,n_0, k)$ with different sparsity parameters $(n_0, k)$; $n_0$ is the initial number of nodes and $k$ is the number of edges a new node attaches to existing nodes.}
\label{fig:MSE_BA}
\end{figure}

Figures \ref{fig:MSE_ER} and \ref{fig:MSE_BA} summarize the MSE of the two methods. These figures show that the proposed approach outperforms the HT estimator for all scenarios in both networks. The MSE converges at the same rate across all scenarios, indicating that the proposed approach is as efficient as the frequentist estimator.  The proposed approach consistently performs well, even under scenarios of model misspecifications (scenarios 2 to 5), demonstrating its efficiency and robustness. 
In particular, the successful results in Scenario 2 highlight that the specific choice of the location function $\mu_k$ is less critical, as discussed in \S\ref{sec:models}. This is attributed to the nonparametric nature of the DPMM, represented by the infinite mixture of distributions characterized by different atoms, which compensates for possibly incorrect specification of $\mu_k$.
In the challenging scenario, Scenario 6, which involves an unmeasured covariate, the MSE slightly increases for both the HT estimator and the proposed method. However, the proposed method continues to demonstrate superior performance, highlighting the reliability of the DoI framework in practical contexts where the data may not be sufficiently rich.
Another interesting observation is that when we have higher-dimensional covariates in Scenario 7, the improvement of our methodology over the traditional HT estimator becomes more pronounced. This is because, in high-dimensional settings, the increased auxiliary information enables our model-assisted approach to capture more of the outcome variation and thus reduce variance more effectively, whereas HT estimator does not fully exploit such auxiliary information.

Further evaluation of bias and coverage for the treatment effects and all metrics for the spillover effects are provided in the Appendix. Those results also demonstrate the efficiency of the proposed method under interference. 


\section{Case Studies}
\label{sec:case_studies_main}
\subsection{Cash Transfer Programs in Colombia}
\label{sec:case_study}
The methodology developed in \S\ref{sec:framework} is applied to a randomized experiment investigating the impact of a conditional cash transfer program, treatment, on students' attendance rates \citep{Barrera-Osorio2011}. The study was conducted in two regions of Bogota, Colombia: San Cristobal and Suba. Households with school children, ranging from one to five children per household, were recruited for the experiment. Within each household, school children were randomly assigned to either enroll in the cash transfer program or not, by a stratified randomization approach based on locality (San Cristobal/Suba),  school type (public/private), gender, and grade level.

Multiple children within a household had the potential to receive the treatment, and within each stratum, all children had an equal chance of being treated. Additionally, given the known treatment assignment, \citet{Barrera-Osorio2011} examined both the direct effect of the program and the possible spillover effect on siblings within the same household. Enrolled students were eligible for cash subsidies if they attended school at least $80\%$ of the time in a given month. Enrolling one student in the program could influence the attendance rate of their sibling(s) in the same household, either positively or negatively. Our objective is to employ the proposed methodology to analyze these direct and spillover effects in the two regions of Bogota, Colombia.

Formally, for student $i$, we denote $Y_{i} \in \{0, 1\}$ where $Y_{i}=1$ if the student is eligible for cash subsidies, i.e., their attendance rate is over $80\%$. We also let $Z_i$ denote a treatment indicator that equals $1$ if student $i$ was enrolled in the program, and $0$ otherwise. We consider the interference occurs within the same household and thus denote $A_{i,j}$ as the adjacency that equals 1 if student $i$ and $j$ are in the same household (siblings), and $0$ otherwise. $\mathcal{N}_i$ denotes the number of siblings of unit $i$. Additionally, $\mathbf{X}_i$ represents the pre-treatment covariates, including the student's age, grade level, gender, household head's age, single-parent household indicator, household size, household's poverty score, household's income status, locality, and the number of students in the household who participated in the lottery. 
% Our analysis focuses on households with complete data and more than one child.
We establish three cluster types based on the two regions and household sizes: households in Suba with two students, households in San Cristobal with two students, and households in San Cristobal with three or more students. Overall, our analysis comprises $1012$ households containing $2135$ students.

We adopt the original treatment randomization probability as the treatment allocation strategy, indicated by the treatment allocation probabilities $\pi =(\pi_{\mathrm{SC}},\pi_{\mathrm{Su}}) = (0.628, 0.449)$ for the two localities San Cristobal and Suba respectively. The primary estimands of interest are E-ATE $\tau_{\text{E-ATE}}(1;\pi)$ and E-ASE $\tau_{\text{E-ASE}}(0;\pi)$ for the current treatment allocation strategy $\pi$, see \S\ref{sec:causal_estimands}.  
%The treatment allocation for the regions of San Cristobal and Suba are denoted as $\pi_{\mathrm{SC}}$ and $\pi_{\mathrm{Su}}$. 
In the given context, the spillover effect measures the discrepancy in the attendance rate of an unenrolled student under the allocation strategy $\pi$ when none of their siblings is enrolled.

Following the discussion in \S\ref{sec:structural_assumption}, we use a probit model
$Y_i(z) \sim \mathrm{Bernoulli} \left(\Psi(\mathbf{X}_i^\top \boldsymbol{\beta}_z +  G_i(\mathbf{T}_i, \mathbf{Z}_{-i}))\right)$ for the $Y$-model, where $\Psi(\cdot)$ denotes the standard normal cumulative distribution function.  The details of the Gibbs sampler for the binary outcome are provided in the Appendix. We here customize the components of $G_i$ to reflect our knowledge about the interference: $\mu(\mathbf{T}_{i}, \Z_{-i}, \gamma_k)= \gamma_{k,1} \sum_{j \in \mathcal{N}_i} Z_jA_{ij} / | \mathcal{N}_i |   + \gamma_{k,2} \left( \text{Grade}_{i} -  \sum_{j \in \mathcal{N}_i} \text{Grade}_{j} Z_jA_{ij} / | \mathcal{N}_i | \right)$, where $\text{Grade}_i$  denotes the grade  of student $i$.
The intuition of this specification is that students are more encouraged to enroll in the program if more or fewer siblings are treated, and they are likely influenced if their siblings are close in grade. 
We use the same proper and weakly informative priors, $\boldsymbol{\beta}_{z} \sim \mathrm{N}((0,0)^\top, 10^2\mathbf{I}_2)$, where $\mathbf{I}_2$ is the $2 \times 2$ identity matrix, $\gamma_{k,1}, \gamma_{k,2} \sim \text{N}(0,10^2)$ and $\sigma_k^2 \sim \text{IG}(0.01,0.01)$. We present a sensitivity analysis with respect to different prior specifications in the following section.

Table \ref{tab:EATE_EASE} presents the estimates of E-ATE and E-ASE under the current treatment allocation strategy $\pi$. We also present the conditional E-ASE for those with $1$ or $2$ siblings and those whose siblings' treated proportion is $33\%$, $50\%$, or $66\%$. We observe that the median of E-ATE is positive and those of E-ASE are mostly negative, implying the potential positive direct treatment effect and negative spillover effects. The spillover effects tend to be slightly larger as the number of siblings or the proportion of treated siblings increases.
Table \ref{tab:EASE_different_strategies} presents the estimates of the E-ASE under different treatment allocation strategies $\pi' \in \{(0.1,0.1),(0.3,0.3),(0.5,0.5),(0.7,0.7),(0.9,0.9)\}$, $\tau_{\text{E-ASE}}(0;\pi')$. We observe that the effect is always negative. As the proportion of treated units increases, the  effects slightly get larger. 

Finally, when performing Bayesian analyses with weakly identifiable models, it is important to investigate the robustness of the results with respect to the prior specifications so as to make inferences more reliable. Considering different prior specifications, we examined the robustness of our analyses; the results are provided in the Appendix. 

\begin{table*}%[htbp]
	\centering
	\caption{Causal estimands and their estimates of posterior means, standard deviation, percentiles, and interval length under the current treatment allocation strategy $\pi$. The superscript $_{nb,k}$ denotes the spillover effects on students with $k$ siblings. The superscript $_{rt,p}$ denotes the spillover effects on students for whom $p\%$ of their siblings are treated. E-ATE, expected average treatment effects; E-ASE, expected average spillover effects. }
	\begin{adjustbox}{width=8.cm}
		\begin{tabular}{lrrrrrr}
			\toprule
                Estimands & Mean & SD & $2.5\%$ & Median & $97.5 \%$ & Length  \\
                \hline
			E-ATE$_{\pi}$ & $ 0.011$ & $0.007$ & $-0.003$ & $0.011$ & $0.025$ & $0.028$ \\
E-ASE$_{\pi}$ & $ -0.003$ & $0.006$ & $-0.014$ & $-0.003$ & $0.009$ & $0.023$ \\
E-ASE$_{\pi}^{nb,1}$ & $ -0.003$ & $0.006$ & $-0.014$ & $-0.003$ & $0.01$ & $0.024$ \\
E-ASE$_{\pi}^{nb,2}$ & $ 0.0$ & $0.017$ & $-0.033$ & $0.0$ & $0.036$ & $0.069$ \\
E-ASE$_{\pi}^{rt,33}$ & $ -0.001$ & $0.026$ & $-0.053$ & $0.0$ & $0.051$ & $0.105$ \\
E-ASE$_{\pi}^{rt,50}$ & $ -0.0$ & $0.008$ & $-0.016$ & $-0.001$ & $0.016$ & $0.032$ \\
E-ASE$_{\pi}^{rt,66}$ & $ 0.004$ & $0.029$ & $-0.051$ & $0.0$ & $0.061$ & $0.113$ \\
			\bottomrule
		\end{tabular}
	\end{adjustbox}
	\label{tab:EATE_EASE}
% \begin{tabnote}
% E-ATE, expected average treatment effects; E-ASE, expected average spillover effects. \raggedright
% \end{tabnote}
\end{table*}

\begin{table*}%[htbp]
	\centering
	\caption{The spillover effects and their estimates of posterior means, standard deviation, percentiles, and interval length under different treatment allocation strategies $\pi'$. E-ASE$_{p}$ denotes expected average spillover effects under the treatment allocation strategy $\pi' = (p,p)$. }
	\begin{adjustbox}{width=8.cm}
		\begin{tabular}{lrrrrrr}
			\toprule
			Estimands & Mean & SD & $2.5\%$ & Median & $97.5 \%$ & Length  \\
			 \hline
          E-ASE$_{\pi}$ & $ -0.003$ & $0.006$ & $-0.014$ & $-0.003$ & $0.009$ & $0.023$ \\
E-ASE$_{0.1}$ & $ -0.011$ & $0.008$ & $-0.029$ & $-0.011$ & $0.004$ & $0.033$ \\
E-ASE$_{0.3}$ & $ -0.009$ & $0.007$ & $-0.023$ & $-0.008$ & $0.005$ & $0.029$ \\
E-ASE$_{0.5}$ & $ -0.006$ & $0.007$ & $-0.019$ & $-0.007$ & $0.006$ & $0.025$ \\
E-ASE$_{0.7}$ & $ -0.004$ & $0.006$ & $-0.017$ & $-0.004$ & $0.008$ & $0.025$ \\
E-ASE$_{0.9}$ & $ -0.002$ & $0.006$ & $-0.015$ & $-0.002$ & $0.01$ & $0.024$ \\
			\bottomrule

		\end{tabular}
	\end{adjustbox}
	\label{tab:EASE_different_strategies}
\end{table*}

\subsection{Inference on the Degree of Interference in an Experiment on Stereolithography}
\label{sec:case_study_3dprinting}

\begin{figure}
\centering
\includegraphics[scale=0.45]{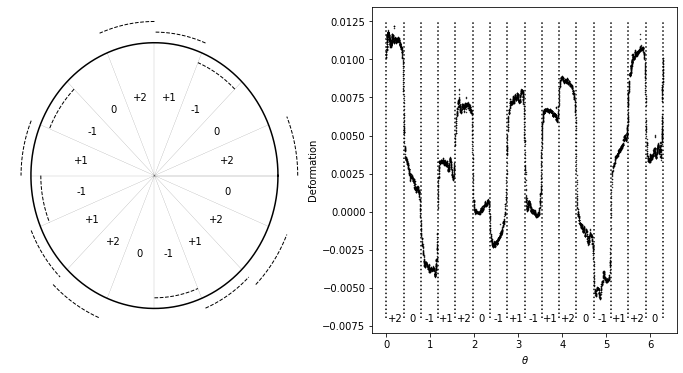}
\caption{Experimental design (left) and observed deformation (right). Dashed lines on the left figure represent assigned compensations. }
\label{fig:experimentaldesign}
\end{figure}

Additive manufacturing (AM), or three-dimensional (3D) printing, refers to a manufacturing technology in which material is deposited and solidified additively based on a computer-aided design (CAD) model to manufacture the corresponding physical product. 
% This technology has many advantages over traditional subtractive manufacturing paradigms. For example, it can reduce the waste associated with subtractive manufacturing, and can enable printing of products with novel geometries and complex inner structures in a straightforward manner \citep{Gibson2009}. 
A significant challenge with AM technologies is dimensional accuracy control of a printed product. Shape deviations inevitably arise in additively manufactured products due to the physics associated with the rapid heating and cooling in any AM process. 
% Shape deviations are defined as the differences between the nominal product (as specified in the CAD model) and the actual printed product \citep{Sabbaghi2014,Huang2015}. 
Several studies have been performed to investigate different strategies for dimensional accuracy control. One strategy is to implement a compensation plan for shape deviations, which corresponds to a modification of the CAD model such that when the modified model is manufactured the corresponding physical product should exhibit reduced shape deviations compared to the original printed product \citep{Huang2015}. 

\citet{Sabbaghi2014} studied shape deviations in a stereolithography process. 
Their experiment considered in-plane deviations for the top layers of four circular cylinders with negligible heights. The cylinder was of nominal radius $0.5''$.
The objective of their experiment was to determine whether interference would arise in the shape deviations for the points on the cylinders under discretized compensation plans. The experimental units in this context are the individual points on the top boundary of a printed product. The covariates for each unit correspond to the location of the unit on a cylinder under the polar coordinate system (i.e., the angle $\psi_i \in (0, 2\pi)$ in radians). The treatment is a continuous treatment factor, corresponding to the amount of compensation (i.e., the addition or subtraction of material) applied to an experimental unit. Each of the potential outcomes $Y_i(\Z)$ for unit $i$ is defined as the difference between the potential and nominal radius at $\psi_i$ under compensation plan $\Z$ applied to the product containing unit $i$. The experimental design corresponds to a restricted Latin square design on a cylinder. The cylinder was divided into $16$ equal-sized sections of $\pi/8$ radians, and one of four levels of the continuous treatment factor was applied to each, with the levels set based on the size of the cylinder. The compensations for the $0.5''$ cylinder were $-0.004'', 0'', 0.004''$, and $0.008''$. The left panel of Figure \ref{fig:experimentaldesign} summarizes the experimental design for the $0.5''$ cylinder. The right panel of this figure summarizes the shape deviation data of all the measured points.

\begin{figure}
\centering
\includegraphics[scale=0.35]{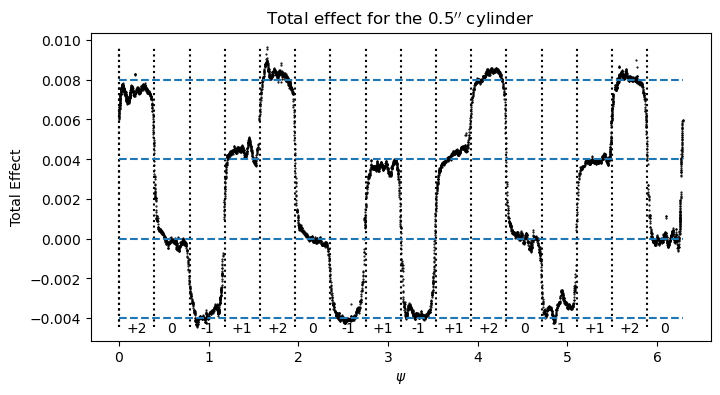}
\includegraphics[scale=0.35]{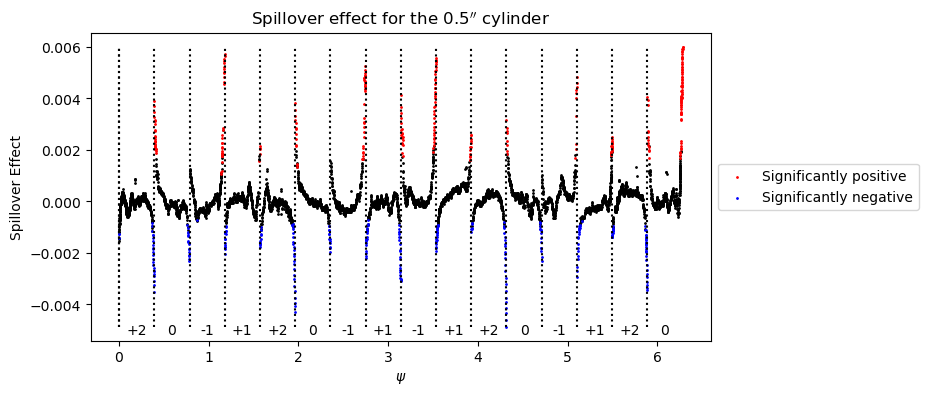}
\caption{Summary of the total effects (left) and spillover effects (right) for all points on the $0.5''$ cylinder as inferred under our DoI framework and Bayesian methodology.}
\label{fig:totaleffect_spillovereffect}
\end{figure}

We proceed to apply our DoI framework to analyze the $N = 6159$ datapoints from the cylinder with nominal radius $r=0.5''$. This application is challenging in the sense that we are interested in the unit-level effects. We first estimate the unit-level assignment-conditional total effects. These are similar in spirit to the effective treatments considered by \citet{Sabbaghi2014}. However, a major difference is that \citet{Sabbaghi2014} defined the estimand in terms of the parameters of their Bayesian model whereas we define the estimand according to the finite-sample perspective, i.e., in terms of observed and missing potential outcomes such as $\tau_i^{\mathrm{tot}}(\z) = Y_i(z_i, \z_{-i})-Y_i(0, \mathbf{0})$. The left panel of Figure \ref{fig:totaleffect_spillovereffect} summarizes the inferences on the total effects obtained from our methodology. These inferences match the effective treatments presented in Figure 5 of \citep{Sabbaghi2014}. Considering the fact that their analysis was based on parametric Bayesian models tailored to this particular AM data set, we conclude that our DoI framework and Bayesian approach successfully captures the effective treatment of compensation as the total effect of compensation in a conceptually more straightforward and general manner. In addition, our DDPM is much more flexible than the parametric models considered by \citet{Sabbaghi2014}, with the flexibility arising from the infinite mixture of kernels with atoms that depend on $\A$. Instead of the common choice of the adjacency matrix, we employed an inverse-distance matrix that was defined as
\begin{equation*}
A_{ij} = 
\begin{cases}
\frac{1}{|\psi_i - \psi_j|} &  \text{ if $|\psi_i - \psi_j| \leq \frac{\pi}{8}$}\\
0 & \text{ otherwise.}
\end{cases}
\end{equation*} 
\noindent The set of neighboring units for unit $i$ is defined as $\mathcal{N}_i = \{j : A_{ij}\neq 0\}$. Our potential outcome model was specified according to $Y_i(z) = \mathbf{X}_i^\top\boldsymbol{\beta}_{Z_i} +  G_i(\mathbf{T}_i, \mathbf{Z}_{-i})$, with $\mu(\mathbf{X}_i, \mathbf{A}_i, \Z_{-i}, \gamma_k) = \gamma_k \sum_{j \in \mathcal{N}_i} Z_jA_{ij}$. Following \citet{Sabbaghi2014} we set $\mathbf{X}_i = (1, \cos\psi_i)^\top$ and specify our prior distributions for the model parameters as $\boldsymbol{\beta}_{Z_i} \sim \mathrm{N}((0,0)^\top, 10\mathbf{I}_2)$, where $\mathbf{I}_2$ is the $2 \times 2$ identity matrix, for $Z_i \in \{-0.004'', 0, 0.004'', 0.008''\}$, $\gamma_k \sim \mathrm{N}(0,10), \sigma_k^2 \sim \text{IG}(1,10^{-7})$, $\sigma^2 \sim \text{IG}(1,10^{-7})$.

In addition to the difference in perspectives, our approach possesses additional advantages. First and foremost, we need only use a single data set to estimate the total effect, whereas \citet{Sabbaghi2014} used two distinct sets of experimental data to assess which units on the cylinders have negligible interference. In particular, their first data set consisted of control data from previously manufactured cylinders for which every unit received zero compensation. They computed the posterior distribution of the model parameters and formed the posterior predictive distribution of the potential outcomes based on these control data. After performing these inferences, they then used the data from the compensation experiment to assess interference in terms of how the units' effective treatments differed from that to which they were physically assigned. This approach is similar to that of \citet{Hudgens2008} in the sense that they prepared multiple clusters of units by their designs and assessed interference by comparing the outcomes in different clusters. In our approach, we only need the data from the compensated products, and not control data on products that did not have a compensation plan, to infer the effects of interference. 

As discussed in \S\ref{sec:causal_estimands}, the unit-level spillover effect is $\tau_i^{\mathrm{sp}}(Z_i, \Z_{-i}, \Z_{-i}') = Y_i(Z_i, \Z_{-i}) - Y_i(Z_i, \Z_{-i}')$ for two different assignment vectors $\Z$ and $\Z^{'}$ with $Z_i = Z_i'$. Following the definition of negligible interference by \citet[p.~1403, 1405]{Sabbaghi2014}, we consider spillover effects in which, for a given $\Z$, we set $\Z' = Z_{i}\mathbf{1}_{N-1}$, where $\mathbf{1}_{N-1}$ is the $(N-1)$-dimensional vector of ones. Our unit-level spillover estimand in this case is formally expressed as $\tau_i^{\mathrm{sp}}(Z_i, \Z_{-i}, \Z_{-i}') = Y_i(Z_i, \Z_{-i}) - Y_i(Z_i, Z_{i}\mathbf{1}_{N-1})$. The right panel of Figure \ref{fig:totaleffect_spillovereffect} presents the spillover effect of all units as inferred from our Bayesian method. The central $95\%$ posterior intervals for significantly positive (negative) units are always positive (negative), and never contain zero, which indicates that these units receive nonzero spillover effects. We observe that our Bayesian method indicates that approximately $90\%$ of units, primarily in the central regions of the $16$ sections of the cylinders, exhibit negligible interference. This result is consistent with that obtained by \citet{Sabbaghi2014}, and was obtained in a more general and straightforward manner compared to \citet{Sabbaghi2014}.

% \begin{figure}
% \centering
% \includegraphics[scale=0.5]{spillovereffects.png}
% \caption{Summary of the spillover effects for all points on the $0.5''$ cylinder as inferred under our DoI framework and Bayesian methodology.}
% \label{fig:spillovereffect}
% \end{figure}

\begin{figure}
\centering
\includegraphics[scale=0.6]{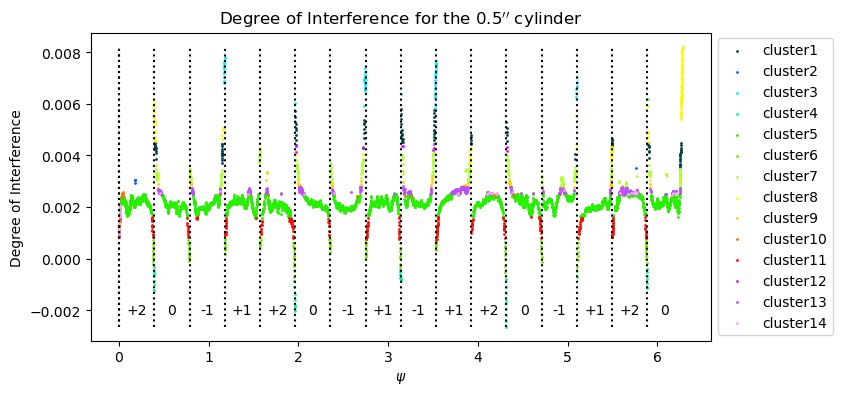}
\caption{Summary of the DoIs for all points on the $0.5''$ cylinder.}
\label{fig:DoI}
\end{figure}

Another advantage of our Bayesian method compared to the work of \citet{Sabbaghi2014} is the clustering property of the DDPM. Figure \ref{fig:DoI} summarizes the DoI of all experimental units as inferred from our model, with different colors representing different clusters of units. The clusters of units are determined by the posterior modes of the latent class indicators $C_i$. Different clusters exhibit different interference structures, which are governed by the atoms of the DDPM model. As seen in this figure, most of the units in the middle of the sections belong to one major cluster (Cluster 6, indicated as green). Units on the edges of the sections belong to other minor clusters and exhibit more significant interference, which corresponds to the fact that in this experiment these units should have different interference structures compared to that of the major cluster. 

Each atom of the DDPM model has its own covariate-dependent location parameter, so that each component corresponds to a single parametric model with a different location parameter. As we obtain multiple clusters, it is clear that modeling the interference structure with a single parametric model, as was done by \citet{Sabbaghi2014}, is not appropriate. Our methodology automatically captures the interference structure and separates the units into different clusters based on the interference effects that they receive. Of the total $N = 6159$ units, $4458$ (or approximately $70 \%$) belong to the major cluster. Clusters $3$, $4$ and $8$ appear only on the edge of the sections in which the absolute difference in their compensation level versus that of the neighboring section is $0.008''$. As such, units in clusters $3$, $4$ and $8$ are highly affected by other units through interference. Finally, we see that the cluster membership changes depending on the distance of a unit from the edge of the section, which gradates the interference structures within sections.

\section{Concluding Remarks}
\label{sec:conclusions}

This article proposed our new notion of the Degree of Interference (DoI) that flexibly handles interference. We presented several new types of causal estimands and developed a fast data augmentation based Bayesian semiparametric methodology for estimating them. In simulation studies across different scenarios with complicated interference structures our approach exhibited better finite-sample performance than that of the conventional HT estimator in terms of MSE. The results provided in Appendix \ref{sec:simulation_supp} also highlight the superiority of our methodology in terms of bias and coverage probability. %We also illustrated our methodology via the analyses of cash transfer programs in Colombia. We also utilize our DoI framework to study shape deviations in a stereolithography process, with details provided in  the Appendix. 
Our two case studies demonstrate that our DoI framework and Bayesian methodology yield more insights and knowledge on interference in a straightforward manner.

Our methodology should complement existing methodologies rather than replace them. For instance, while our approach introduces modeling assumptions, existing methods \citep{Fredrik2021_2, hoshino2023} are fully nonparametric, which may appeal to analysts reluctant to make modeling assumptions. Conversely, analysts may find the conservative variance estimate provided by nonparametric methods too large and seek efficiency gains through our proposed method.

A possible future direction to build upon this framework would be to extend our framework to observational studies where the treatment assignment mechanisms take unknown functional forms.
% \yo{Additionally, when the dimensionality of covariates becomes ultrahigh (e.g., $ d > N $), both computational and statistical challenges arise, as is the case with many traditional statistical techniques. For instance, in our Gibbs sampler, we compute the inverse of $ X^\top X $, which becomes singular (non-invertible) in such cases, resulting in unstable estimates with high variance. extending our methodology to handle high-dimensional cases as a promising direction for future research.}
% Also, experimental designs to efficiently learn the DoI would be another future research topic.

\appendix

\section{Detailed Steps of Gibbs Sampler}
\label{sec:detail_Gibbs}
In this section we present the detailed steps of the Gibbs sampler developed in \S\ref{sec:Gibbssampling_outline}. First, our implementation of the Gibbs sampler is outlined below.
    \begin{enumerate}
		\item Initialize parameters $\boldsymbol{\phi}^{(0)}$, $\boldsymbol{\theta}^{(0)}$.
		
		\item Repeat the following steps. For $t = 0, 1, \ldots$
		
		\begin{enumerate}
			
			\item Draw $\mathbf{G}^{\mathrm{o}, (t+1)} \sim p \left ( \mathbf{G} \mid \mathbf{T}, \mathbf{Y}^{\mathrm{obs}}, \boldsymbol{\phi}^{(t)},  \boldsymbol{\theta}^{(t)} \right )$.
			
			\item Draw $\left ( \boldsymbol{\phi}^{(t+1)}, \boldsymbol{\theta}^{(t+1)} \right ) \sim p \left ( \boldsymbol{\phi}, \boldsymbol{\theta} \mid  \mathbf{T},  \mathbf{Y}^{\mathrm{obs}},
			\mathbf{G}^{\mathrm{o}, (t+1)}  \right )$.
			
			\item Draw $\mathbf{G}^{\mathrm{u}, (t+1)} \sim p \left ( \mathbf{G} \mid \mathbf{T},  \mathbf{Y}^{\mathrm{obs}}, \mathbf{G}^{\mathrm{o}, (t+1)}, \boldsymbol{\phi}^{(t+1)} , \boldsymbol{\theta}^{(t+1)} \right )$.
			
			\item Draw $\mathbf{Y}^{\mathrm{mis}, (t+1)} \sim p \left ( \mathbf{Y}^{\mathrm{mis}} \mid \mathbf{T},  \mathbf{Y}^{\mathrm{obs}}, \mathbf{G}^{\mathrm{o}, (t+1)}, \mathbf{G}^{\mathrm{u}, (t+1)}, \boldsymbol{\phi}^{(t+1)}, \boldsymbol{\theta}^{(t+1)}  \right )$.
			
			\item Calculate the causal estimand based on the observed and imputed data.
			
			\end{enumerate}
	\end{enumerate}
\noindent 
The detailed steps are provided below.
\begin{enumerate}
    \item  For the units with $C_i=k$ and $Z_i=z$, draw $G_i^o$ from $\mathrm{N}\big(\{\lambda_z(\gamma_0 + \gamma_k \sum_{j \in \mathcal{N}_i} Z_jA_{ij})+(Y_i-X_i\beta_z)\sigma_k^2\}/(\lambda_z+\sigma_k^2), \lambda_z\sigma_k^2/(\lambda_z+\sigma_k^2)\big)$
    % \begin{align*}
    %     \mathrm{N}\bigg(\{\lambda_z(\gamma_0 + \gamma_k \sum_{j \in \mathcal{N}_i} Z_jA_{ij})+(Y_i-X_i\beta_z)\sigma_k^2\}/(\lambda_z+\sigma_k^2), \lambda_z\sigma_k^2/(\lambda_z+\sigma_k^2)\bigg)
    % \end{align*}
    \item For each unit, draw $C_i$ from $\pr(C_i=k|-) = \frac{w_k \mathrm{N}(G_i|\gamma_0 + \gamma_k \sum_{j \in \mathcal{N}_i} Z_jA_{ij}, \sigma_k^2)}{\sum_{k=1}^{K}w_k \mathrm{N}(G_i|\gamma_0 + \gamma_k \sum_{j \in \mathcal{N}_i} Z_jA_{ij}, \sigma_k^2)} $
    % \begin{align*}
    %     \pr(C_i=k|-) = \frac{w_k \mathrm{N}(G_i|\gamma_0 + \gamma_k \sum_{j \in \mathcal{N}_i} Z_jA_{ij}, \sigma_k^2)}{\sum_{k=1}^{K}w_k \mathrm{N}(G_i|\gamma_0 + \gamma_k \sum_{j \in \mathcal{N}_i} Z_jA_{ij}, \sigma_k^2)} 
    % \end{align*}
    \item Update $w_k$'s as Step 3 in \S\ref{sec:Gibbssampling_outline}.
    \item We adopt the Metropolis Hasting step with a proposal draw $\alpha^{*} \sim \mathrm{Gamma(1,1)}$. Given $\mathbf{w}'$ and $\mathbf{C}$, we accept the draw $\alpha^{*}$ with probability $r$ where
    \begin{align*}
        r = \min\bigg\{1, \frac{p \left( \alpha^{*} \mid 1,1 \right) \prod_{k=1}^{K} f \left(w_k' \mid 1+ \sum_{i:C_i=k}1, \alpha^{*}+ \sum_{i:C_i>k}1 \right)}{p \left( \alpha \mid 1,1 \right) \prod_{k=1}^{K} f \left(w_k' \mid 1+ \sum_{i:C_i=k}1, \alpha+ \sum_{i:C_i>k}1 \right)} \bigg\}.
    \end{align*}
     We reject the draw $\alpha^{*}$ with probability $1-r$ and accept the draw $\alpha$, obtained in the previous iteration. Note that $p \left( \alpha \mid a,b \right)$ denotes the prior distribution of $\alpha$, which we specify by the Gamma distribution with parameters $a=1,b=1$ and $f$ is specified in  Step 4 in \S\ref{sec:Gibbssampling_outline}.
    \item 
    \begin{enumerate}
        \item If $N_k=\sum_{i=1}^{N}\mathbbm{1}(C_i=k)>0$, draw $\sigma_k^2$ from $\mathrm{IG}(2+0.5N_k, 1+0.5s_k)$ where $s_k=\sum_{i:C_i=k}(G_i^o - \gamma_0 - \gamma_k \sum_{j \in \mathcal{N}_i} Z_jA_{ij})^2$. If $N_k=0$,then draw $\sigma_k^2$ from the prior $\mathrm{IG}(2,1)$.
        \item If $N_k>0$, draw $\gamma_k$ from $\mathrm{N}\big((\mathbf{M_k}^\top\mathbf{M_k}+\sigma_k^2/10 \times I_2)^{-1} \mathbf{M_k}^\top \mathbf{G}_{k}^o, \sigma_k^2 (\mathbf{M_k}^\top\mathbf{M_k}+\sigma_k^2/10 \times I_2)^{-1}\big)$
        % \begin{align*}
        %     \mathrm{N}\bigg((\mathbf{M_k}^\top\mathbf{M_k}+\sigma_k^2/10 \times I_2)^{-1} \mathbf{M_k}^\top \mathbf{G}_{k}^o, \sigma_k^2 (\mathbf{M_k}^\top\mathbf{M_k}+\sigma_k^2/10 \times I_2)^{-1}\bigg)
        % \end{align*}
        where $\mathbf{G}_k^o$ is a subvector of $\mathbf{G}^o$ for units with $C_i=k$ and $\mathbf{M_k}$ is an $N_k \times 2$ matrix with the row vector $(1, \sum_{j \in \mathcal{N}_i} Z_jA_{ij})$ for units with $C_i=k$. If $N_k=0$, then draw from $\mathrm{N}((0,0)^\top, \sigma_k^2/10 \times I_2)$.
    \end{enumerate}
    \item 
    \begin{enumerate}
        \item Draw $\lambda_z$ from $\mathrm{IG}(2+N_z/2, 1+0.5\sum_{i:Z_i=z}(Y_i - X_i\beta_z -G_i^o)^2)$.
        \item Draw $\beta_z$ from $\mathrm{N}((\mathbf{X}_z^\top\mathbf{X}_z + \lambda_z/10\times I_2)^{-1}\mathbf{X}_z^\top(\mathbf{Y}_z - \mathbf{G}_z^o), \lambda_z(\mathbf{X}_z^\top\mathbf{X}_z + \lambda_z/10\times I_2)^{-1})$.
        where $\mathbf{G}_z^o)$, $\mathbf{X}_z$ and $\mathbf{Y}_z$ are respectively a submatrix and a subvector of $\mathbf{G}^o$, $\mathbf{X}$ and $\mathbf{Y}$ for units with $Z_i=z$. 
    \end{enumerate}
    \item [7-10.] Follow the procedures in \S\ref{sec:Gibbssampling_outline} for assignment-conditional effects or the steps provided in the next section for the expected effects, depending on which of the effects (conditional effects or expected effects) one is interested in.
\end{enumerate}

\subsection{Estimation of Expected Effects}
\label{sec:Gibbs_expectedeffects}

We present an extension to infer the expected effects. 
It suffices to derive the procedures for estimating $\E_{\z_{-i} \sim \pi}[Y_i(z, \z_{-i})]$ where $\pi$ is the marginalized distribution of $\Z_{-i}$, which depends on the experimental design. 
% The sampling of the marginals $\Z_{-i}$ are achieved through the samples from the joint distribution, i.e., we can sample $\Z \sim \pi$ and just throw away $Z_i$ for the marginal of $\Z_{-i}$. 
For notational convenience, the outer integral below is taken with respect to the counting measure $\pi$ for the distribution of $Z_{-i}$.
\begin{equation*}
    \begin{split}
        \E_{\z_{-i} \sim \pi}[Y_i(z, \z_{-i})] 
        &= \E_{\z_{-i} \sim \pi}\bigg[Y_i(z, \z_{-i}) \int_{\mathcal{G}}dF_{G_i|\Z_{-i}=\z_{-i}}(g)\bigg]
        = \E_{\z_{-i} \sim \pi}\bigg[\int_{\mathcal{G}} Y_i(z, \z_{-i}) dF_{G_i|\Z_{-i}=\z_{-i}}(g)\bigg]\\
        &= \int_{\Omega^{N-1}}\int_{\mathcal{G}} Y_i(z, g) dF_{G_i|\Z_{-i}=\z_{-i}}(g) d\pi(\z_{-i})
          \approx \frac{1}{M} \sum_{m=1}^{M} Y_i(z, g_i^{(m)})
    \end{split}
\end{equation*}
$g_i^{(m)}$ is obtained by the following two steps. For $m=1,...,M$, (1) $\z^{m} \sim \pi$, and then (2) $g_i^{(m)} \sim F_{G_i|\Z_{-i}=\z_{-i}^{m}}(g)$
where $\z^{m}_{-i}$ is a subvector of $\z^{m}$ with the $i$-th entry removed. 
We infer the expected effects by following the same steps from step 1 to step 6 in \S\ref{sec:Gibbssampling_outline}, and then we proceed as:
\begin{enumerate}\setcounter{enumi}{6}
    \item Draw $\z^{*} \sim \pi$.
    \item Given $\mathbf{w}$ in step 3, draw $C_i^{*}$ from $C_i^{*} \sim \text{MN}(\mathbf{w})$.
    \item 
    \begin{enumerate}
        \item Given $C_i^{*}$, $\boldsymbol{\gamma}$ and $\boldsymbol{\sigma}^2$, draw $G_i^{*}$ from $G_i^{*} \sim \text{N}(\mu(\mathbf{T}_i,\z_{-i}^{*},\gamma_{C_i^{*}}),\sigma_{C_i^{*}}^2)$.
        \item Given $C_i^{*}$, $\boldsymbol{\gamma}$ and $\boldsymbol{\sigma}^2$, draw $G_i^{0}$ from $G_i^{0} \sim \text{N}(\mu(\mathbf{T}_i, \mathbf{0}^{N-1},\gamma_{C_i^{*}}),\sigma_{C_i^{*}}^2)$.
    \end{enumerate}
    \item 
    \begin{enumerate}
        \item Given $G_i^{*}$ in step 9(a) and $\theta$ in step 6, draw from $Y_i(z, \z_{-i}^{*}) \sim \pr(Y_i | Z_i=z,\mathbf{T}_i,G_i^{*},\theta)$ for $z=0,1$.
        \item Given $G_i^{0}$ in step 9(b) and $\theta$ in step 6, draw from $Y_i(0, \mathbf{0}^{N-1}) \sim \pr(Y_i | Z_i=0,\mathbf{T}_i,G_i^{0},\theta)$.
    \end{enumerate}
\end{enumerate}

\subsection{Extention to Binary Outcomes}
When we adopt a Probit model as in our illustrative analyses in \S\ref{sec:case_study}, we draw the sample of  $\beta_z$ as follows.
Given $G_{i}^{o} = g_i$ and $\beta_{z}$ in the previous iteration,  we first draw an auxiliary variable $v_{i}$ for unit $i$ with $Z_{i}=z$ from
\begin{align*}
    v_{i} \sim \begin{cases}
    \mathrm{TN}(X_i \beta_{z} + g_i, 1, 0, \infty) \text{ if } Y_{i} = 1, \\
    \mathrm{TN}(X_i \beta_{z} + g_i, 1, -\infty, 0 ) \text{ if } Y_{i} = 0,
    \end{cases}
\end{align*}
where $\mathrm{TN}(\mu,\sigma^2,l,u)$ denotes the truncated normal distribution with the mean, variance, lower bound, and upper bound parameters.
Then, we draw from $\beta_{z} \sim \mathrm{N} (M, V)$,
where $V = \left( H_0^{-1} + \mathbf{X}_z^\top \mathbf{X}_z\right)^{-1}, M = V(H_0^{-1}h_0 + \mathbf{X}_z^\top\mathbf{v}_{z})$
with $\mathbf{v}_{z}$  being a subvector of $\mathbf{v}=(v_1,\ldots,v_N)$ for units with $Z_{i}=z$. 
The prior distribution of $\beta_{z}$ is specified here by $\beta_{z}\sim \mathrm{N}\left(h_0,H_0 \right)$.

\section{Simulation Results}
\label{sec:simulation_supp}
This section presents the simulation evaluation not presented in \S\ref{sec:simulation_studies}; bias and coverage for E-ATE, and bias, MSE, and coverage for E-ASE. We consider the same data-generating mechanisms as in \S\ref{sec:simulation_studies}.
Table \ref{tab:simulation_EASE_1} - \ref{tab:simulation_EASE_6} present the results.

For E-ATE, the DoI approach outperforms the HT estimator in terms of MSE and bias for almost all scenarios of both networks, even under severe model misspecifications (Scenario 2 -- 5), and in the presence of an unmeasured covariate. The MSE decreases as $N$ increases, and the coverage probability is well-calibrated with $95\%$ probability. These observations demonstrate the efficiency and robustness of our DoI approach. 
The HT estimator sometimes exhibits large bias when $N$ is small, whereas the Bayesian methodology consistently works well. Both methodologies exhibit smaller bias as $N$ increases. They are well-calibrated in that they yield intervals with nearly $95\%$ coverage across nearly all conditions for the E-ATE, indicating the robustness of our methodologies for misspecifications. 
Another advantage of the DoI methodology is that it provides a principled estimation procedure for the spillover effects. Overall, the DoI methodology yields good bias, MSE, and coverage for the E-ASE across all scenarios. 
% Only when the data-generating process is scenario 4, it sometimes exhibits slightly larger bias and smaller coverages, especially when the model is severely misspecified (Scenario 4, ER($p=0.03, 0.05$)). This phenomenon can be explained in terms of the connectedness of the underlying graph. 
% The Erd\"os-R\'enyi (ER) graph is known to be almost surely connected when $p > \frac{(1+\epsilon) \log N}{N}$ for any $\epsilon>0$, and disconnected otherwise \citep{frieze_karonski_2015}. Therefore, when $p=0.05$ and $N=1000$, the ER graph is likely to be fully connected, hence the influence of misspecification inflates and dominates the flexibility of our models; that is, in the case of a misspecified model, the Bayesian posterior credible intervals could be discrepant with the confidence interval. In contrast, the BA graph usually yields disconnected graphs with multiple subgraphs from its construction, leading to the good performance of our methodology for the E-ASE under the BA graphs. All scenarios exhibit good performance when the underlying graph is the Barab\'asi–Albert (BA) graph. This observation should have minimal impact on our empirical analyses because any of our datasets in the empirical analyses do not have fully connected graphs as the underlying structure, and have multiple independent connected subgraphs in it. 

\begin{table*}
	\centering
	\caption{Evaluation metrics of simulations studies (Scenario 1). The data-generating process (DGP) and the underlying network between units are provided in \S\ref{sec:simulation_studies}.}
	\begin{adjustbox}{width=15cm}
		\begin{tabular}{lllrrrrrrrrr}
			\toprule
			& & & \multicolumn{3}{c}{HT (E-ATE)}&  \multicolumn{3}{c}{DoI (E-ATE)}&  \multicolumn{3}{c}{DoI (E-ASE)} \\
              \cmidrule(lr){4-6} \cmidrule(lr){7-9} \cmidrule(lr){10-12}
			DGP & Network (parameters) & $N$ & Bias & MSE & Coverage & Bias & MSE & Coverage& Bias & MSE & Coverage  \\
			 \hline
			Scenario 1 & ER$(p=0.01)$ & $300$ & $-0.0061$ & $0.0513$ & $0.99$ & $0.002$ & $0.0136$ & $0.97$ & $-0.0013$ & $0.0074$ & $0.97$ \\
  &  & $500$ & $-0.0055$ & $0.0321$ & $0.975$ & $-0.0018$ & $0.0086$ & $0.95$ & $0.0019$ & $0.0077$ & $0.99$ \\
  &  & $1000$ & $-0.0113$ & $0.0156$ & $0.98$ & $-0.0032$ & $0.0033$ & $0.99$ & $0.0018$ & $0.0068$ & $0.995$ \\
 & ER$(p=0.03)$ & $300$ & $-0.0052$ & $0.0515$ & $0.99$ & $0.0025$ & $0.0133$ & $0.965$ & $-0.0133$ & $0.0205$ & $0.98$ \\
 &  & $500$ & $-0.0064$ & $0.0319$ & $0.98$ & $-0.0017$ & $0.0088$ & $0.96$ & $-0.0096$ & $0.019$ & $0.975$ \\
 &  & $1000$ & $-0.0117$ & $0.0155$ & $0.98$ & $-0.0029$ & $0.0033$ & $0.995$ & $-0.0144$ & $0.0242$ & $0.965$ \\
 & ER$(p=0.05)$ & $300$ & $-0.005$ & $0.0512$ & $0.99$ & $0.0028$ & $0.0134$ & $0.955$ & $-0.0037$ & $0.0347$ & $0.975$ \\
 &  & $500$ & $-0.0057$ & $0.032$ & $0.98$ & $-0.0018$ & $0.0087$ & $0.955$ & $-0.0377$ & $0.034$ & $0.985$ \\
 &  & $1000$ & $-0.0115$ & $0.0155$ & $0.98$ & $-0.0027$ & $0.0033$ & $0.995$ & $-0.0179$ & $0.0338$ & $0.975$ \\
 & BA$(n_0=10,k=3)$ & $300$ & $-0.0056$ & $0.0522$ & $0.99$ & $0.0027$ & $0.0133$ & $0.975$ & $-0.0051$ & $0.0098$ & $0.98$ \\
 &  & $500$ & $-0.0061$ & $0.0318$ & $0.98$ & $-0.0016$ & $0.0087$ & $0.95$ & $-0.0104$ & $0.0071$ & $0.985$ \\
 & & $1000$ & $-0.0116$ & $0.0154$ & $0.98$ & $-0.0026$ & $0.0033$ & $0.99$ & $-0.0035$ & $0.0039$ & $0.97$ \\
 & BA$(n_0=10,k=5)$ & $300$ & $-0.0058$ & $0.0513$ & $0.995$ & $0.0009$ & $0.0134$ & $0.965$ & $-0.0285$ & $0.0185$ & $0.975$ \\
 &  & $500$ & $-0.0065$ & $0.0316$ & $0.98$ & $-0.0017$ & $0.0087$ & $0.955$ & $-0.0056$ & $0.0129$ & $0.985$ \\
 & & $1000$ & $-0.0116$ & $0.0155$ & $0.98$ & $-0.003$ & $0.0033$ & $0.99$ & $-0.0088$ & $0.0074$ & $0.945$ \\
 & BA$(n_0=30,k=3)$ & $300$ & $-0.0052$ & $0.0513$ & $0.99$ & $0.0022$ & $0.0136$ & $0.96$ & $-0.0182$ & $0.0104$ & $0.965$ \\
 &  & $500$ & $-0.0059$ & $0.0322$ & $0.975$ & $-0.0022$ & $0.0087$ & $0.95$ & $-0.0075$ & $0.006$ & $0.975$ \\
 & & $1000$ & $-0.0115$ & $0.0155$ & $0.98$ & $-0.0026$ & $0.0034$ & $0.99$ & $-0.0054$ & $0.0042$ & $0.95$ \\
 & BA$(n_0=30,k=5)$ & $300$ & $-0.0045$ & $0.0511$ & $0.99$ & $0.0021$ & $0.0134$ & $0.96$ & $-0.005$ & $0.0115$ & $0.985$ \\
 &  & $500$ & $-0.006$ & $0.0318$ & $0.98$ & $-0.0013$ & $0.0087$ & $0.955$ & $0.0012$ & $0.0078$ & $0.975$ \\
 &  & $1000$ & $-0.0118$ & $0.0155$ & $0.98$ & $-0.003$ & $0.0033$ & $0.995$ & $0.0006$ & $0.007$ & $0.945$ \\
			\bottomrule
		\end{tabular}
	\end{adjustbox}
	\label{tab:simulation_EASE_1}
\end{table*}

\begin{table*}
	\centering
	\caption{Evaluation metrics of simulations studies  (Scenario 2). }
	\begin{adjustbox}{width=15cm}
		\begin{tabular}{lllrrrrrrrrr}
			\toprule
			& & & \multicolumn{3}{c}{HT (E-ATE)}&  \multicolumn{3}{c}{DoI (E-ATE)}&  \multicolumn{3}{c}{DoI (E-ASE)} \\
              \cmidrule(lr){4-6} \cmidrule(lr){7-9} \cmidrule(lr){10-12}
			DGP & Network (parameters) & $N$ & Bias & MSE & Coverage & Bias & MSE & Coverage& Bias & MSE & Coverage  \\
			 \hline
			Scenario 2 & ER$(p=0.01)$ & $300$ & $-0.0047$ & $0.0508$ & $0.99$ & $0.0031$ & $0.0134$ & $0.97$ & $0.0044$ & $0.0074$ & $0.96$ \\
  &  & $500$ & $-0.0058$ & $0.0309$ & $0.98$ & $-0.0015$ & $0.0086$ & $0.955$ & $0.0053$ & $0.0077$ & $0.985$ \\
  &  & $1000$ & $-0.0106$ & $0.0152$ & $0.98$ & $-0.0026$ & $0.0033$ & $0.995$ & $0.005$ & $0.0069$ & $0.995$ \\
 & ER$(p=0.03)$ & $300$ & $-0.0046$ & $0.051$ & $0.99$ & $0.0019$ & $0.0131$ & $0.965$ & $-0.0067$ & $0.0209$ & $0.98$ \\
 &  & $500$ & $-0.0059$ & $0.0311$ & $0.98$ & $-0.0005$ & $0.0087$ & $0.95$ & $-0.0021$ & $0.0198$ & $0.985$ \\
 &  & $1000$ & $-0.0111$ & $0.0152$ & $0.98$ & $-0.0025$ & $0.0033$ & $0.995$ & $-0.0079$ & $0.0248$ & $0.95$ \\
 & ER$(p=0.05)$ & $300$ & $-0.0048$ & $0.051$ & $0.99$ & $0.003$ & $0.0134$ & $0.955$ & $0.0054$ & $0.035$ & $0.975$ \\
 &  & $500$ & $-0.0058$ & $0.0312$ & $0.98$ & $-0.0018$ & $0.0085$ & $0.95$ & $-0.0275$ & $0.0337$ & $0.985$ \\
 &  & $1000$ & $-0.0111$ & $0.0153$ & $0.98$ & $-0.003$ & $0.0033$ & $0.99$ & $-0.0098$ & $0.0347$ & $0.965$ \\
 & BA$(n_0=10,k=3)$ & $300$ & $-0.0048$ & $0.0511$ & $0.99$ & $0.0022$ & $0.0131$ & $0.97$ & $0.0001$ & $0.01$ & $0.98$ \\
 &  & $500$ & $-0.006$ & $0.0305$ & $0.98$ & $-0.0014$ & $0.0086$ & $0.95$ & $-0.0071$ & $0.007$ & $0.98$ \\
 & & $1000$ & $-0.0109$ & $0.0151$ & $0.98$ & $-0.003$ & $0.0033$ & $0.985$ & $-0.0008$ & $0.004$ & $0.97$ \\
 & BA$(n_0=10,k=5)$ & $300$ & $-0.0052$ & $0.051$ & $0.99$ & $0.0007$ & $0.0134$ & $0.96$ & $-0.0204$ & $0.0182$ & $0.975$ \\
 &  & $500$ & $-0.0062$ & $0.0304$ & $0.98$ & $-0.0015$ & $0.0086$ & $0.955$ & $-0.0011$ & $0.013$ & $0.98$ \\
 & & $1000$ & $-0.0109$ & $0.0151$ & $0.98$ & $-0.003$ & $0.0033$ & $0.995$ & $-0.0061$ & $0.0075$ & $0.95$ \\
 & BA$(n_0=30,k=3)$ & $300$ & $-0.0049$ & $0.051$ & $0.99$ & $0.0032$ & $0.0136$ & $0.955$ & $-0.0136$ & $0.0103$ & $0.97$ \\
 &  & $500$ & $-0.0061$ & $0.0305$ & $0.98$ & $-0.0011$ & $0.0086$ & $0.95$ & $-0.0046$ & $0.006$ & $0.975$ \\
 & & $1000$ & $-0.0108$ & $0.015$ & $0.98$ & $-0.0027$ & $0.0033$ & $0.99$ & $-0.0031$ & $0.0042$ & $0.945$ \\
 & BA$(n_0=30,k=5)$ & $300$ & $-0.0048$ & $0.0509$ & $0.99$ & $0.0027$ & $0.0132$ & $0.965$ & $0.0009$ & $0.0116$ & $0.98$ \\
 &  & $500$ & $-0.006$ & $0.0305$ & $0.98$ & $-0.0016$ & $0.0088$ & $0.955$ & $0.0058$ & $0.0079$ & $0.985$ \\
 &  & $1000$ & $-0.0109$ & $0.0151$ & $0.98$ & $-0.0023$ & $0.0032$ & $0.995$ & $0.0032$ & $0.0071$ & $0.925$ \\
			\bottomrule
		\end{tabular}
	\end{adjustbox}
	\label{tab:simulation_EASE_2}
\end{table*}

\begin{table*}
	\centering
	\caption{Evaluation metrics of simulations studies  (Scenario 3). }
	\begin{adjustbox}{width=15cm}
		\begin{tabular}{lllrrrrrrrrr}
			\toprule
			& & & \multicolumn{3}{c}{HT (E-ATE)}&  \multicolumn{3}{c}{DoI (E-ATE)}&  \multicolumn{3}{c}{DoI (E-ASE)} \\
              \cmidrule(lr){4-6} \cmidrule(lr){7-9} \cmidrule(lr){10-12}
			DGP & Network (parameters) & $N$ & Bias & MSE & Coverage & Bias & MSE & Coverage& Bias & MSE & Coverage  \\
			 \hline
			Scenario 3 & ER$(p=0.01)$ & $300$ & $0.0033$ & $0.0678$ & $0.91$ & $0.0034$ & $0.0132$ & $0.975$ & $0.0049$ & $0.0074$ & $0.97$ \\
  &  & $500$ & $0.0048$ & $0.0333$ & $0.945$ & $-0.0017$ & $0.0086$ & $0.96$ & $0.0089$ & $0.0078$ & $0.975$ \\
  &  & $1000$ & $-0.0029$ & $0.0168$ & $0.95$ & $-0.0022$ & $0.0033$ & $0.99$ & $0.01$ & $0.0069$ & $0.995$ \\
 & ER$(p=0.03)$ & $300$ & $0.0034$ & $0.066$ & $0.915$ & $0.0038$ & $0.0133$ & $0.965$ & $-0.0017$ & $0.021$ & $0.975$ \\
 &  & $500$ & $0.0045$ & $0.0329$ & $0.945$ & $-0.0018$ & $0.0088$ & $0.95$ & $0.0031$ & $0.0194$ & $0.975$ \\
 &  & $1000$ & $-0.0035$ & $0.0167$ & $0.955$ & $-0.0024$ & $0.0033$ & $0.99$ & $-0.0036$ & $0.0246$ & $0.95$ \\
 & ER$(p=0.05)$ & $300$ & $0.0032$ & $0.066$ & $0.915$ & $0.0027$ & $0.0133$ & $0.965$ & $0.0095$ & $0.0349$ & $0.975$ \\
 &  & $500$ & $0.0046$ & $0.0328$ & $0.945$ & $-0.0013$ & $0.0087$ & $0.95$ & $-0.0223$ & $0.0335$ & $0.99$ \\
 &  & $1000$ & $-0.0035$ & $0.0166$ & $0.95$ & $-0.0031$ & $0.0033$ & $0.995$ & $-0.0045$ & $0.0347$ & $0.965$ \\
 & BA$(n_0=10,k=3)$ & $300$ & $0.0032$ & $0.0675$ & $0.91$ & $0.0029$ & $0.013$ & $0.975$ & $0.003$ & $0.0099$ & $0.98$ \\
 &  & $500$ & $0.0044$ & $0.0336$ & $0.94$ & $-0.0017$ & $0.0087$ & $0.95$ & $-0.0051$ & $0.007$ & $0.975$ \\
 & & $1000$ & $-0.0034$ & $0.0173$ & $0.955$ & $-0.0021$ & $0.0033$ & $0.995$ & $0.0004$ & $0.004$ & $0.97$ \\
 & BA$(n_0=10,k=5)$ & $300$ & $0.0027$ & $0.0676$ & $0.915$ & $0.0014$ & $0.0133$ & $0.975$ & $-0.0188$ & $0.0183$ & $0.98$ \\
 &  & $500$ & $0.0042$ & $0.0336$ & $0.94$ & $-0.0017$ & $0.0088$ & $0.95$ & $0.0011$ & $0.013$ & $0.975$ \\
 & & $1000$ & $-0.0034$ & $0.0173$ & $0.955$ & $-0.0025$ & $0.0032$ & $0.995$ & $-0.0041$ & $0.0075$ & $0.94$ \\
 & BA$(n_0=30,k=3)$ & $300$ & $0.0031$ & $0.0675$ & $0.91$ & $0.0039$ & $0.0136$ & $0.955$ & $-0.0111$ & $0.0102$ & $0.965$ \\
 &  & $500$ & $0.0043$ & $0.0336$ & $0.945$ & $-0.0012$ & $0.0087$ & $0.95$ & $-0.0025$ & $0.006$ & $0.965$ \\
 & & $1000$ & $-0.0033$ & $0.0172$ & $0.955$ & $-0.0026$ & $0.0033$ & $0.99$ & $-0.0016$ & $0.0042$ & $0.95$ \\
 & BA$(n_0=30,k=5)$ & $300$ & $0.0031$ & $0.0675$ & $0.91$ & $0.0021$ & $0.0131$ & $0.97$ & $0.0027$ & $0.0115$ & $0.98$ \\
 &  & $500$ & $0.0044$ & $0.0336$ & $0.94$ & $-0.001$ & $0.0087$ & $0.955$ & $0.0074$ & $0.008$ & $0.985$ \\
 &  & $1000$ & $-0.0033$ & $0.0172$ & $0.955$ & $-0.003$ & $0.0033$ & $0.995$ & $0.0052$ & $0.0071$ & $0.925$ \\
			\bottomrule
		\end{tabular}
	\end{adjustbox}
	\label{tab:simulation_EASE_3}
\end{table*}

\begin{table*}
	\centering
	\caption{Evaluation metrics of simulations studies  (Scenario 4). }
	\begin{adjustbox}{width=15cm}
		\begin{tabular}{lllrrrrrrrrr}
			\toprule
			& & & \multicolumn{3}{c}{HT (E-ATE)}&  \multicolumn{3}{c}{DoI (E-ATE)}&  \multicolumn{3}{c}{DoI (E-ASE)} \\
              \cmidrule(lr){4-6} \cmidrule(lr){7-9} \cmidrule(lr){10-12}
			DGP & Network (parameters) & $N$ & Bias & MSE & Coverage & Bias & MSE & Coverage& Bias & MSE & Coverage  \\
			 \hline
			Scenario 4 & ER$(p=0.01)$ & $300$ & $-0.0049$ & $0.0508$ & $0.99$ & $0.0034$ & $0.0133$ & $0.97$ & $0.0007$ & $0.0074$ & $0.96$ \\
  &  & $500$ & $-0.0062$ & $0.0302$ & $0.98$ & $-0.0016$ & $0.0087$ & $0.95$ & $-0.0064$ & $0.0078$ & $0.98$ \\
  &  & $1000$ & $-0.0114$ & $0.0152$ & $0.975$ & $-0.0032$ & $0.0034$ & $0.99$ & $-0.0129$ & $0.0069$ & $0.985$ \\
 & ER$(p=0.03)$ & $300$ & $-0.0056$ & $0.0507$ & $0.985$ & $0.0011$ & $0.0136$ & $0.955$ & $-0.0235$ & $0.0212$ & $0.98$ \\
 &  & $500$ & $-0.007$ & $0.0298$ & $0.98$ & $-0.0021$ & $0.009$ & $0.95$ & $-0.0192$ & $0.0194$ & $0.975$ \\
 &  & $1000$ & $-0.0117$ & $0.0152$ & $0.975$ & $-0.0033$ & $0.0035$ & $0.99$ & $-0.0311$ & $0.0252$ & $0.965$ \\
 & ER$(p=0.05)$ & $300$ & $-0.0053$ & $0.0503$ & $0.99$ & $0.0026$ & $0.0137$ & $0.965$ & $-0.0128$ & $0.0346$ & $0.965$ \\
 &  & $500$ & $-0.0072$ & $0.0298$ & $0.98$ & $-0.0025$ & $0.009$ & $0.95$ & $-0.048$ & $0.0353$ & $0.98$ \\
 &  & $1000$ & $-0.0121$ & $0.0154$ & $0.975$ & $-0.0036$ & $0.0036$ & $0.99$ & $-0.0357$ & $0.0349$ & $0.975$ \\
 & BA$(n_0=10,k=3)$ & $300$ & $-0.0053$ & $0.0509$ & $0.99$ & $0.0019$ & $0.0133$ & $0.97$ & $-0.0061$ & $0.0098$ & $0.98$ \\
 &  & $500$ & $-0.0063$ & $0.0301$ & $0.98$ & $-0.0004$ & $0.0086$ & $0.95$ & $-0.0137$ & $0.0071$ & $0.98$ \\
 & & $1000$ & $-0.0111$ & $0.0151$ & $0.98$ & $-0.0027$ & $0.0033$ & $0.995$ & $-0.0054$ & $0.004$ & $0.965$ \\
 & BA$(n_0=10,k=5)$ & $300$ & $-0.0055$ & $0.0506$ & $0.99$ & $0.0007$ & $0.0133$ & $0.975$ & $-0.0271$ & $0.0187$ & $0.975$ \\
 &  & $500$ & $-0.0065$ & $0.0301$ & $0.98$ & $-0.0026$ & $0.0086$ & $0.945$ & $-0.0072$ & $0.0129$ & $0.985$ \\
 & & $1000$ & $-0.0112$ & $0.0151$ & $0.98$ & $-0.0029$ & $0.0034$ & $0.995$ & $-0.0099$ & $0.0074$ & $0.95$ \\
 & BA$(n_0=30,k=3)$ & $300$ & $-0.0051$ & $0.0509$ & $0.985$ & $0.003$ & $0.0136$ & $0.965$ & $-0.0208$ & $0.0106$ & $0.965$ \\
 &  & $500$ & $-0.0063$ & $0.0302$ & $0.98$ & $-0.0011$ & $0.0087$ & $0.945$ & $-0.0105$ & $0.0061$ & $0.965$ \\
 & & $1000$ & $-0.0108$ & $0.015$ & $0.98$ & $-0.0029$ & $0.0033$ & $0.995$ & $-0.007$ & $0.0043$ & $0.95$ \\
 & BA$(n_0=30,k=5)$ & $300$ & $-0.005$ & $0.0507$ & $0.99$ & $0.0021$ & $0.0132$ & $0.97$ & $-0.0065$ & $0.0116$ & $0.985$ \\
 &  & $500$ & $-0.0063$ & $0.0302$ & $0.98$ & $-0.002$ & $0.0088$ & $0.955$ & $-0.0001$ & $0.008$ & $0.98$ \\
 &  & $1000$ & $-0.0111$ & $0.0151$ & $0.98$ & $-0.0034$ & $0.0034$ & $0.995$ & $-0.0008$ & $0.007$ & $0.94$ \\
			\bottomrule
		\end{tabular}
	\end{adjustbox}
	\label{tab:simulation_EASE_4}
\end{table*}

\begin{table*}
	\centering
	\caption{Evaluation metrics of simulations studies  (Scenario 5). }
	\begin{adjustbox}{width=15cm}
		\begin{tabular}{lllrrrrrrrrr}
			\toprule
			& & & \multicolumn{3}{c}{HT (E-ATE)}&  \multicolumn{3}{c}{DoI (E-ATE)}&  \multicolumn{3}{c}{DoI (E-ASE)} \\
              \cmidrule(lr){4-6} \cmidrule(lr){7-9} \cmidrule(lr){10-12}
			DGP & Network (parameters) & $N$ & Bias & MSE & Coverage & Bias & MSE & Coverage& Bias & MSE & Coverage  \\
			 \hline
			Scenario 5 & ER$(p=0.01)$ & $300$ & $-0.0047$ & $0.0508$ & $0.99$ & $0.0029$ & $0.013$ & $0.97$ & $0.0041$ & $0.0073$ & $0.97$ \\
  &  & $500$ & $-0.0058$ & $0.0308$ & $0.98$ & $-0.0011$ & $0.0086$ & $0.955$ & $0.0059$ & $0.0078$ & $0.98$ \\
  &  & $1000$ & $-0.0107$ & $0.0152$ & $0.98$ & $-0.0027$ & $0.0032$ & $0.99$ & $0.0036$ & $0.0068$ & $0.99$ \\
 & ER$(p=0.03)$ & $300$ & $-0.0048$ & $0.0511$ & $0.99$ & $0.0023$ & $0.0133$ & $0.97$ & $-0.0067$ & $0.0207$ & $0.98$ \\
 &  & $500$ & $-0.0063$ & $0.0308$ & $0.98$ & $-0.0018$ & $0.0087$ & $0.95$ & $-0.0027$ & $0.0193$ & $0.97$ \\
 &  & $1000$ & $-0.0113$ & $0.0153$ & $0.98$ & $-0.0035$ & $0.0033$ & $0.985$ & $-0.0088$ & $0.0244$ & $0.96$ \\
 & ER$(p=0.05)$ & $300$ & $-0.0049$ & $0.0511$ & $0.99$ & $0.0026$ & $0.0133$ & $0.97$ & $0.0056$ & $0.0351$ & $0.975$ \\
 &  & $500$ & $-0.0062$ & $0.0309$ & $0.98$ & $-0.0016$ & $0.0086$ & $0.955$ & $-0.0275$ & $0.0336$ & $0.985$ \\
 &  & $1000$ & $-0.0115$ & $0.0154$ & $0.98$ & $-0.003$ & $0.0033$ & $0.995$ & $-0.0099$ & $0.034$ & $0.97$ \\
 & BA$(n_0=10,k=3)$ & $300$ & $-0.0049$ & $0.0511$ & $0.99$ & $0.0026$ & $0.0132$ & $0.965$ & $0.0008$ & $0.0098$ & $0.98$ \\
 &  & $500$ & $-0.006$ & $0.0305$ & $0.98$ & $-0.0009$ & $0.0086$ & $0.955$ & $-0.0074$ & $0.007$ & $0.985$ \\
 & & $1000$ & $-0.0109$ & $0.0151$ & $0.98$ & $-0.0027$ & $0.0033$ & $0.995$ & $-0.001$ & $0.0039$ & $0.965$ \\
 & BA$(n_0=10,k=5)$ & $300$ & $-0.0052$ & $0.051$ & $0.99$ & $-0.0$ & $0.0134$ & $0.965$ & $-0.0201$ & $0.0182$ & $0.975$ \\
 &  & $500$ & $-0.0063$ & $0.0304$ & $0.98$ & $-0.002$ & $0.0085$ & $0.95$ & $-0.0011$ & $0.0129$ & $0.985$ \\
 & & $1000$ & $-0.011$ & $0.0151$ & $0.98$ & $-0.0032$ & $0.0033$ & $0.99$ & $-0.0063$ & $0.0075$ & $0.945$ \\
 & BA$(n_0=30,k=3)$ & $300$ & $-0.0049$ & $0.051$ & $0.99$ & $0.0032$ & $0.0137$ & $0.96$ & $-0.0131$ & $0.0104$ & $0.975$ \\
 &  & $500$ & $-0.0061$ & $0.0305$ & $0.98$ & $-0.0017$ & $0.0087$ & $0.94$ & $-0.0046$ & $0.006$ & $0.965$ \\
 & & $1000$ & $-0.0108$ & $0.015$ & $0.98$ & $-0.0024$ & $0.0033$ & $0.995$ & $-0.0029$ & $0.0042$ & $0.955$ \\
 & BA$(n_0=30,k=5)$ & $300$ & $-0.0048$ & $0.0509$ & $0.99$ & $0.0019$ & $0.0131$ & $0.965$ & $-0.0008$ & $0.0116$ & $0.985$ \\
 &  & $500$ & $-0.0061$ & $0.0305$ & $0.98$ & $-0.0013$ & $0.0087$ & $0.945$ & $0.0063$ & $0.008$ & $0.98$ \\
 &  & $1000$ & $-0.0109$ & $0.0151$ & $0.98$ & $-0.0032$ & $0.0033$ & $0.98$ & $0.0032$ & $0.007$ & $0.93$ \\
			\bottomrule
		\end{tabular}
	\end{adjustbox}
	\label{tab:simulation_EASE_5}
\end{table*}

\begin{table*}
	\centering
	\caption{Evaluation metrics of simulations studies  (Scenario 6). }
	\begin{adjustbox}{width=15cm}
		\begin{tabular}{lllrrrrrrrrr}
			\toprule
			& & & \multicolumn{3}{c}{HT (E-ATE)}&  \multicolumn{3}{c}{DoI (E-ATE)}&  \multicolumn{3}{c}{DoI (E-ASE)} \\
              \cmidrule(lr){4-6} \cmidrule(lr){7-9} \cmidrule(lr){10-12}
			DGP & Network (parameters) & $N$ & Bias & MSE & Coverage & Bias & MSE & Coverage& Bias & MSE & Coverage  \\
			 \hline
			Scenario 6 & ER$(p=0.01)$ & $300$ & $0.0068$ & $0.0775$ & $0.99$ & $0.008$ & $0.0279$ & $0.99$ & $-0.0062$ & $0.0087$ & $0.985$ \\
  &  & $500$ & $0.0004$ & $0.0468$ & $0.98$ & $-0.0016$ & $0.0172$ & $0.99$ & $0.0039$ & $0.0103$ & $0.98$ \\
  &  & $1000$ & $-0.0099$ & $0.0213$ & $0.975$ & $0.0017$ & $0.0076$ & $0.995$ & $0.0069$ & $0.0132$ & $0.955$ \\
 & ER$(p=0.03)$ & $300$ & $0.0065$ & $0.0777$ & $0.99$ & $0.0053$ & $0.0277$ & $0.99$ & $-0.0128$ & $0.0302$ & $0.985$ \\
 &  & $500$ & $0.0006$ & $0.047$ & $0.99$ & $-0.0018$ & $0.0172$ & $0.99$ & $-0.0035$ & $0.0325$ & $0.97$ \\
 &  & $1000$ & $-0.01$ & $0.0214$ & $0.975$ & $0.0025$ & $0.0076$ & $1.0$ & $-0.0222$ & $0.0339$ & $0.975$ \\
 & ER$(p=0.05)$ & $300$ & $0.0068$ & $0.0783$ & $0.99$ & $0.0076$ & $0.0283$ & $0.995$ & $-0.0361$ & $0.0398$ & $0.98$ \\
 &  & $500$ & $0.0007$ & $0.047$ & $0.99$ & $-0.0009$ & $0.0176$ & $0.99$ & $-0.0221$ & $0.041$ & $0.985$ \\
 &  & $1000$ & $-0.01$ & $0.0214$ & $0.975$ & $0.0025$ & $0.0077$ & $1.0$ & $-0.0293$ & $0.0547$ & $0.965$ \\
 & BA$(n_0=10,k=3)$ & $300$ & $0.0066$ & $0.0779$ & $0.99$ & $0.0088$ & $0.0279$ & $1.0$ & $-0.0034$ & $0.0182$ & $0.97$ \\
 &  & $500$ & $0.001$ & $0.0466$ & $0.99$ & $-0.0014$ & $0.0176$ & $0.99$ & $0.0095$ & $0.0116$ & $0.97$ \\
 & & $1000$ & $-0.0098$ & $0.021$ & $0.975$ & $0.0016$ & $0.0076$ & $1.0$ & $0.0013$ & $0.0065$ & $0.97$ \\
 & BA$(n_0=10,k=5)$ & $300$ & $0.0068$ & $0.078$ & $0.99$ & $0.0077$ & $0.028$ & $0.99$ & $-0.0236$ & $0.0265$ & $0.975$ \\
 &  & $500$ & $0.0012$ & $0.0469$ & $0.99$ & $-0.0005$ & $0.0173$ & $0.99$ & $-0.0016$ & $0.0222$ & $0.95$ \\
 & & $1000$ & $-0.0092$ & $0.0212$ & $0.975$ & $0.0026$ & $0.0076$ & $1.0$ & $0.0007$ & $0.0094$ & $0.99$ \\
 & BA$(n_0=30,k=3)$ & $300$ & $0.0068$ & $0.0779$ & $0.99$ & $0.0074$ & $0.0283$ & $0.985$ & $-0.0057$ & $0.0124$ & $0.98$ \\
 &  & $500$ & $0.0011$ & $0.0466$ & $0.985$ & $-0.0019$ & $0.0176$ & $0.99$ & $-0.0117$ & $0.0092$ & $0.975$ \\
 & & $1000$ & $-0.0095$ & $0.0211$ & $0.975$ & $0.0023$ & $0.0077$ & $1.0$ & $-0.0043$ & $0.0062$ & $0.97$ \\
 & BA$(n_0=30,k=5)$ & $300$ & $0.0071$ & $0.0772$ & $0.99$ & $0.0066$ & $0.0282$ & $0.985$ & $-0.0079$ & $0.0192$ & $0.975$ \\
 &  & $500$ & $0.0011$ & $0.0468$ & $0.985$ & $-0.0023$ & $0.0174$ & $0.99$ & $-0.0217$ & $0.014$ & $0.98$ \\
 &  & $1000$ & $-0.0098$ & $0.0213$ & $0.975$ & $0.002$ & $0.0077$ & $1.0$ & $-0.0054$ & $0.0086$ & $0.97$ \\
			\bottomrule
		\end{tabular}
	\end{adjustbox}
	\label{tab:simulation_EASE_6}
\end{table*}

\begin{table*}
	\centering
	\caption{Evaluation metrics of simulations studies  (Scenario 7). }
	\begin{adjustbox}{width=15cm}
		\begin{tabular}{lllrrrrrrrrr}
			\toprule
			& & & \multicolumn{3}{c}{HT (E-ATE)}&  \multicolumn{3}{c}{DoI (E-ATE)}&  \multicolumn{3}{c}{DoI (E-ASE)} \\
              \cmidrule(lr){4-6} \cmidrule(lr){7-9} \cmidrule(lr){10-12}
			DGP & Network (parameters) & $N$ & Bias & MSE & Coverage & Bias & MSE & Coverage& Bias & MSE & Coverage  \\
			 \hline
			Scenario 7 & ER$(p=0.01)$ & $300$ & $0.0486$ & $0.0958$ & $1.0$ & $0.0074$ & $0.013$ & $0.985$ & $0.0008$ & $0.0064$ & $0.99$ \\
  &  & $500$ & $-0.0267$ & $0.0684$ & $0.96$ & $0.0039$ & $0.0088$ & $0.955$ & $0.0041$ & $0.0069$ & $0.975$ \\
  &  & $1000$ & $0.0105$ & $0.0263$ & $0.995$ & $0.0038$ & $0.0039$ & $0.955$ & $-0.0007$ & $0.0101$ & $0.955$ \\
 & ER$(p=0.03)$ & $300$ & $0.0476$ & $0.0953$ & $1.0$ & $0.0073$ & $0.0129$ & $0.98$ & $-0.0215$ & $0.0229$ & $0.96$ \\
 &  & $500$ & $-0.0269$ & $0.0684$ & $0.96$ & $0.004$ & $0.0089$ & $0.94$ & $-0.0147$ & $0.0198$ & $0.98$ \\
 &  & $1000$ & $0.0105$ & $0.0263$ & $0.995$ & $0.0045$ & $0.0038$ & $0.965$ & $0.0059$ & $0.021$ & $0.975$ \\
 & ER$(p=0.05)$ & $300$ & $0.0471$ & $0.0955$ & $1.0$ & $0.0086$ & $0.0128$ & $0.985$ & $-0.0039$ & $0.0355$ & $0.975$ \\
 &  & $500$ & $-0.027$ & $0.0683$ & $0.96$ & $0.0038$ & $0.009$ & $0.935$ & $-0.0363$ & $0.0338$ & $0.975$ \\
 &  & $1000$ & $0.0108$ & $0.0265$ & $0.995$ & $0.004$ & $0.0038$ & $0.965$ & $-0.015$ & $0.0373$ & $0.965$ \\
 & BA$(n_0=10,k=3)$ & $300$ & $0.0478$ & $0.0952$ & $1.0$ & $0.0082$ & $0.0129$ & $0.985$ & $-0.002$ & $0.011$ & $0.975$ \\
 &  & $500$ & $-0.027$ & $0.0683$ & $0.96$ & $0.0028$ & $0.0088$ & $0.955$ & $-0.0052$ & $0.0071$ & $0.97$ \\
 & & $1000$ & $0.0105$ & $0.0263$ & $0.995$ & $0.0038$ & $0.0039$ & $0.96$ & $0.0062$ & $0.0036$ & $0.975$ \\
 & BA$(n_0=10,k=5)$ & $300$ & $0.0475$ & $0.0956$ & $1.0$ & $0.007$ & $0.0131$ & $0.98$ & $-0.0151$ & $0.0192$ & $0.96$ \\
 &  & $500$ & $-0.0271$ & $0.0685$ & $0.96$ & $0.0033$ & $0.0089$ & $0.945$ & $-0.0054$ & $0.0125$ & $0.975$ \\
 & & $1000$ & $0.0106$ & $0.0264$ & $0.995$ & $0.0049$ & $0.0038$ & $0.965$ & $0.0017$ & $0.0066$ & $0.935$ \\
 & BA$(n_0=30,k=3)$ & $300$ & $0.0485$ & $0.0955$ & $1.0$ & $0.0071$ & $0.0128$ & $0.98$ & $-0.0145$ & $0.0092$ & $0.98$ \\
 &  & $500$ & $-0.0273$ & $0.0681$ & $0.96$ & $0.0034$ & $0.0089$ & $0.945$ & $-0.0075$ & $0.006$ & $0.97$ \\
 & & $1000$ & $0.0105$ & $0.0264$ & $0.995$ & $0.004$ & $0.0038$ & $0.96$ & $-0.0025$ & $0.0036$ & $0.97$ \\
 & BA$(n_0=30,k=5)$ & $300$ & $0.0479$ & $0.0957$ & $1.0$ & $0.0074$ & $0.013$ & $0.985$ & $-0.0068$ & $0.0129$ & $0.955$ \\
 &  & $500$ & $-0.0275$ & $0.0682$ & $0.96$ & $0.0026$ & $0.0089$ & $0.955$ & $-0.008$ & $0.0105$ & $0.955$ \\
 &  & $1000$ & $0.0108$ & $0.0262$ & $0.995$ & $0.0037$ & $0.0038$ & $0.965$ & $-0.0117$ & $0.0062$ & $0.945$ \\
			\bottomrule
		\end{tabular}
	\end{adjustbox}
	\label{tab:simulation_EASE_7}
\end{table*}

\newpage

\section{Sensitivity Analyses for the Case Study in the Main Text}
\label{sec:sensitivity_supp}
Throughout the manuscript, we use proper, although weakly informative, conjugate prior distributions for all parameters. In this section, we examine the robustness of the results concerning the prior specifications. We present sensitivity analyses with the following prior specifications: 
Table \ref{tab:sensitivity2}--\ref{tab:sensitivity4} report the results with the following priors:
\newline \noindent 
\underline{Prior 1:} 
$\boldsymbol{\beta}_{z},  \sim \mathrm{N}((0,0)^\top, 20^2\mathbf{I}_2)$, $\gamma_{k,1}, \gamma_{k,2}  \sim \mathrm{N}(0, 20^2)$, $\sigma_k^2 \sim \text{IG}(0.1,0.1)$.
\newline \noindent 
\underline{Prior 2:} 
$\boldsymbol{\beta}_{z},  \sim \mathrm{N}((0,0)^\top, 5^2\mathbf{I}_2)$, $\gamma_{k,1}, \gamma_{k,2}  \sim \mathrm{N}(0, 5^2)$, $\sigma_k^2 \sim \text{IG}(0.1,0.1)$.
\newline \noindent 
\underline{Prior 3:} 
$\boldsymbol{\beta}_{z}, \sim \mathrm{N}((0,0)^\top, 10^2\mathbf{I}_2)$, $\gamma_{k,1}, \gamma_{k,2}  \sim \mathrm{N}(0, 10^2)$, $\sigma_k^2 \sim \text{IG}(0.01,0.01)$.
\newline \noindent 
\underline{Prior 4:} 
$\boldsymbol{\beta}_{z}, \sim \mathrm{N}((0,0)^\top, 10^2\mathbf{I}_2)$, $\gamma_{k,1}, \gamma_{k,2}  \sim \mathrm{N}(0, 10^2)$, $\sigma_k^2 \sim \text{IG}(1.0,1.0)$. 
\newline \noindent
Compared with Table \ref{tab:EATE_EASE}, the results change only slightly, validating the robustness of our analyses.

\begin{table*}
	\centering
	\caption{Sensitivity Analysis for different prior specifications (Prior 1).}
	\begin{adjustbox}{width=10.cm}
		\begin{tabular}{lrrrrrr}
			\toprule
			Estimands & Mean & SD & $2.5\%$ & Median & $97.5 \%$ & Length  \\
                \hline
			E-ATE$_{\pi}$ & $ 0.011$ & $0.007$ & $-0.002$ & $0.011$ & $0.026$ & $0.028$ \\
E-ASE$_{\pi}$ & $ -0.004$ & $0.006$ & $-0.016$ & $-0.004$ & $0.008$ & $0.024$ \\
E-ASE$_{\pi}^{nb,1}$ & $ -0.004$ & $0.006$ & $-0.016$ & $-0.004$ & $0.009$ & $0.026$ \\
E-ASE$_{\pi}^{nb,2}$ & $ -0.002$ & $0.018$ & $-0.036$ & $-0.004$ & $0.033$ & $0.069$ \\
E-ASE$_{\pi}^{rt,33}$ & $ -0.003$ & $0.027$ & $-0.055$ & $0.0$ & $0.05$ & $0.105$ \\
E-ASE$_{\pi}^{rt,50}$ & $ -0.003$ & $0.008$ & $-0.019$ & $-0.004$ & $0.013$ & $0.033$ \\
E-ASE$_{\pi}^{rt,66}$ & $ -0.0$ & $0.03$ & $-0.058$ & $0.0$ & $0.06$ & $0.118$ \\
			\bottomrule
		\end{tabular}
	\end{adjustbox}
	\label{tab:sensitivity1}
\end{table*}

\begin{table*}
	\centering
	\caption{Sensitivity Analysis for different prior specifications (Prior 2).}
	\begin{adjustbox}{width=10.cm}
		\begin{tabular}{lrrrrrr}
			\toprule
			Estimands & Mean & SD & $2.5\%$ & Median & $97.5 \%$ & Length  \\
			 \hline
          E-ATE$_{\pi}$ & $ 0.01$ & $0.006$ & $-0.002$ & $0.009$ & $0.023$ & $0.025$ \\
E-ASE$_{\pi}$ & $ -0.0$ & $0.006$ & $-0.011$ & $-0.0$ & $0.012$ & $0.023$ \\
E-ASE$_{\pi}^{nb,1}$ & $ -0.0$ & $0.006$ & $-0.012$ & $0.0$ & $0.013$ & $0.024$ \\
E-ASE$_{\pi}^{nb,2}$ & $ 0.002$ & $0.017$ & $-0.033$ & $0.0$ & $0.036$ & $0.069$ \\
E-ASE$_{\pi}^{rt,33}$ & $ 0.002$ & $0.026$ & $-0.048$ & $0.0$ & $0.054$ & $0.101$ \\
E-ASE$_{\pi}^{rt,50}$ & $ 0.0$ & $0.008$ & $-0.015$ & $0.0$ & $0.016$ & $0.031$ \\
E-ASE$_{\pi}^{rt,66}$ & $ 0.003$ & $0.03$ & $-0.052$ & $0.0$ & $0.065$ & $0.117$ \\
\bottomrule
		\end{tabular}
	\end{adjustbox}
	\label{tab:sensitivity2}
\end{table*}

\begin{table*}
	\centering
	\caption{Sensitivity Analysis for different prior specifications (Prior 3).}
	\begin{adjustbox}{width=10.cm}
		\begin{tabular}{lrrrrrr}
			\toprule
			Estimands & Mean & SD & $2.5\%$ & Median & $97.5 \%$ & Length  \\
			 \hline
         E-ATE$_{\pi}$ & $ 0.011$ & $0.007$ & $-0.002$ & $0.011$ & $0.026$ & $0.028$ \\
E-ASE$_{\pi}$ & $ -0.003$ & $0.006$ & $-0.015$ & $-0.003$ & $0.01$ & $0.025$ \\
E-ASE$_{\pi}^{nb,1}$ & $ -0.003$ & $0.007$ & $-0.015$ & $-0.003$ & $0.01$ & $0.025$ \\
E-ASE$_{\pi}^{nb,2}$ & $ -0.0$ & $0.018$ & $-0.033$ & $0.0$ & $0.036$ & $0.069$ \\
E-ASE$_{\pi}^{rt,33}$ & $ -0.001$ & $0.026$ & $-0.052$ & $0.0$ & $0.051$ & $0.103$ \\
E-ASE$_{\pi}^{rt,50}$ & $ -0.002$ & $0.008$ & $-0.017$ & $-0.002$ & $0.015$ & $0.032$ \\
E-ASE$_{\pi}^{rt,66}$ & $ 0.001$ & $0.03$ & $-0.056$ & $0.0$ & $0.06$ & $0.116$ \\
\bottomrule
		\end{tabular}
	\end{adjustbox}
	\label{tab:sensitivity3}
\end{table*}

\begin{table*}
	\centering
	\caption{Sensitivity Analysis for different prior specifications (Prior 4).}
	\begin{adjustbox}{width=10.cm}
		\begin{tabular}{lrrrrrr}
			\toprule
			Estimands & Mean & SD & $2.5\%$ & Median & $97.5 \%$ & Length  \\
			 \hline
          E-ATE$_{\pi}$ & $ 0.012$ & $0.007$ & $-0.001$ & $0.011$ & $0.027$ & $0.028$ \\
E-ASE$_{\pi}$ & $ -0.003$ & $0.006$ & $-0.015$ & $-0.003$ & $0.008$ & $0.024$ \\
E-ASE$_{\pi}^{nb,1}$ & $ -0.003$ & $0.006$ & $-0.016$ & $-0.003$ & $0.009$ & $0.025$ \\
E-ASE$_{\pi}^{nb,2}$ & $ -0.002$ & $0.018$ & $-0.036$ & $-0.004$ & $0.033$ & $0.069$ \\
E-ASE$_{\pi}^{rt,33}$ & $ -0.002$ & $0.027$ & $-0.053$ & $0.0$ & $0.052$ & $0.104$ \\
E-ASE$_{\pi}^{rt,50}$ & $ -0.003$ & $0.008$ & $-0.019$ & $-0.004$ & $0.013$ & $0.032$ \\
E-ASE$_{\pi}^{rt,66}$ & $ -0.001$ & $0.029$ & $-0.058$ & $0.0$ & $0.059$ & $0.117$ \\
\bottomrule
		\end{tabular}
	\end{adjustbox}
	\label{tab:sensitivity4}
\end{table*}

\bibliographystyle{Chicago}
\bibliography{paper-ref}
% \end{singlespace}
\end{document}